\documentclass[useAMS,usenatbib,twocolumn]{mn2e}
\usepackage{amsmath,amssymb}
\usepackage{natbib}
\usepackage[dvipdfmx]{graphicx}
\usepackage{times}
\usepackage{rotate}
\usepackage[colorlinks=true,urlcolor=blue,linkcolor=blue,citecolor=blue]{hyperref}
\usepackage{color,url}
\usepackage{subfigure}

\usepackage{etoolbox}
\makeatletter
\patchcmd{\NAT@citex}
  {\@citea\NAT@hyper@{\NAT@nmfmt{\NAT@nm}\NAT@date}}
  {\@citea\NAT@nmfmt{\NAT@nm}\NAT@hyper@{\NAT@date}}
  {}
  {}
\patchcmd{\NAT@citex}
  {\@citea\NAT@hyper@{%
     \NAT@nmfmt{\NAT@nm}%
     \hyper@natlinkbreak{\NAT@aysep\NAT@spacechar}{\@citeb\@extra@b@citeb}%
     \NAT@date}}
  {\@citea\NAT@nmfmt{\NAT@nm}%
   \NAT@aysep\NAT@spacechar%
   \NAT@hyper@{\NAT@date}}
  {}
  {}
\patchcmd{\NAT@citex}
  {\@citea\NAT@hyper@{%
     \NAT@nmfmt{\NAT@nm}%
     \hyper@natlinkbreak{\NAT@spacechar\NAT@@open\if*#1*\else#1\NAT@spacechar\fi}%
       {\@citeb\@extra@b@citeb}%
     \NAT@date}}
  {\@citea\NAT@nmfmt{\NAT@nm}%
   \NAT@spacechar\NAT@@open\if*#1*\else#1\NAT@spacechar\fi%
   \NAT@hyper@{\NAT@date}}
  {}
  {}
\makeatother

\usepackage{cprotect}
\usepackage{breqn}
\usepackage{xspace}
\usepackage{soul}

\newcommand\sext{\textsc{SExtractor}\xspace}
\newcommand\req{$R_{\rm eq}$ }
\newcommand\Aeff{$A_{\rm effect}$ }

\newcommand{\aap}{A\&A}
\newcommand{\aaps}{A\&AS}
\newcommand{\mnras}{MNRAS}
\newcommand{\apj}{ApJ}

\newcommand{\apjl}{ApJ}

\newcommand{\apjs}{ApJS}

\newcommand{\nat}{Nature}

\newcommand{\pasp}{PASP}

\newcommand{\simgt}{\lower.5ex\hbox{$\; \buildrel > \over \sim \;$}}
\newcommand{\simlt}{\lower.5ex\hbox{$\; \buildrel < \over \sim \;$}}
\newcommand{\ave}[1]{\left\langle #1\right\rangle}

\newcommand{\eg}{{\it e.g.}}

\newcommand{\sn}{S{\rm /}N\xspace}

\newcommand{\msun}{\mbox{${\rm M_{\odot}}$}\xspace}

\newcommand{\mtwob}{M_{200{\rm b}}^{}}

\begin{document}


 \title[Weak lensing on 20 kiloparsec scales]{Can we use Weak Lensing to Measure Total Mass Profiles of Galaxies on 20 Kiloparsec Scales?}


 \author[Kobayashi et al.]
        {Masato I.N. Kobayashi$^1$,
         Alexie Leauthaud$^2$,
         Surhud More$^2$,
         Nobuhiro Okabe$^2$,
         \and
         Clotilde Laigle$^{3,4}$,
         Jason Rhodes$^{5,6}$,
         and
         Tsutomu T. Takeuchi$^1$\\
         $^1$
         Division of Particle and Astrophysical Science, Graduate School of Science, Nagoya University, Aichi 464-8602, Japan\\
         $^2$
         Kavli Institute for the Physics and Mathematics of the Universe (Kavli IPMU, WPI), The University of Tokyo, Chiba 277-8582, Japan\\
         $^3$
         CNRS, UMR 7095, Institut d’Astrophysique de Paris, 98 bis Boulevard Arago, F-75014 Paris, France\\
         $^4$
         Sorbonne Universités, UPMC Univ Paris 06, UMR 7095, Institut d’Astrophysique de Paris, F-75005 Paris, France\\
         $^5$
         Jet Propulsion Laboratory, California Institute of Technology,Pasadena, CA 91109, United States of America\\
         $^6$
         California Institute of Technology, Pasadena, CA 91125, United States of America\\
         }

\maketitle
\label{firstpage}


\begin{abstract}
  Current constraints on dark matter density profiles from weak
  lensing are typically limited to radial scales greater than
  $50$--$100$ kpc. In this paper, we explore the possibility of
  probing the very inner regions of galaxy/halo density profiles by
  measuring stacked weak lensing on scales of only a few tens of
  kpc. Our forecasts focus on scales smaller than the ``equality
  radius" ($R_{\rm eq}$) where the stellar component and the dark
  matter component contribute equally to the lensing signal. We
  compute the evolution of $R_{\rm eq}$ as a function of lens stellar
  mass and redshift and show that $R_{\rm eq}=7$--$34$ kpc for
  galaxies with $M_*=10^{9.5}$--$10^{11.5}$\msun. Unbiased shear
  measurements will be challenging on these scales. We introduce a
  simple metric to quantify how many source galaxies overlap with
  their neighbours and for which shear measurements will be
  challenging. Rejecting source galaxies with close-by companions
  results in a $\sim20$ per cent decrease in the overall source
  density. Despite this decrease, we show that {\it Euclid} and {\it
    WFIRST} will be able to constrain galaxy/halo density profiles
  at $R_{\rm eq}$ with \sn $>20$ for $M_*>10^{10}$\msun. Weak lensing
  measurements at $R_{\rm eq}$, in combination with stellar kinematics
  on smaller scales, will be a powerful means by which to constrain
  both the inner slope of the dark matter density profile as well as
  the mass and redshift dependence of the stellar initial mass
  function. \end{abstract}





\section{Introduction}
\label{sec:introduction}

According to the current cosmological framework, structure formation
in the Universe is driven by the dynamics of cold dark matter.
The collisionless gravitational collapse of dark matter overdensities
and their subsequent virialization leads to the formation of dark
matter halos with different masses and sizes.  A variety of large
scale cold dark matter numerical simulations
\citep[\eg,][]{Dubinski1991,Navarro1996} have shown that the mass
density distribution within halos is generally well described by a one
parameter family, commonly referred to as the Navarro, Frenk, \& White
density profile (\citealt[]{Navarro1997}, hereafter NFW profile).

The power-law slope of the dark matter distribution on scales of a few
kpc to a few tens of kpc can provide clues to the nature of dark
matter (\citealt[]{Flores1994,Moore1994,Weinberg2013}).  In
particular, models with significant warm dark matter
\citep[\eg,][]{Maccio2012} or large self-interaction cross-sections
(\citealt[]{Spergel2000,Peter2013,Zavala2013}) can result in a
shallower density distribution in the core.  However, baryonic physics
also can significantly alter the distribution of dark matter on small
scales, either by feedback
(\citealt[]{Navarro1996b,Mashchenko2006,Zolotov2012,Arraki2014}) or
simply by gravitational effects (adiabatic contraction:
\citealt[]{Blumenthal1986,Gnedin2004,Sellwood2005}).

On radial scales below about one effective radius, the total mass
profiles of galaxies transition from a dark matter dominated regime to a
star-dominated regime. In addition, gas may represent a significant
contribution in low mass galaxies. Disentangling the dark matter
component from the stellar component on these scales is
challenging. There are significant systematic uncertainties in the
determination of galaxy stellar masses from the integrated light
coming out of stars. Variations in the stellar initial mass
function (hereafter IMF) and the low mass cut-off for star formation
result in a factor of two uncertainty in stellar mass estimates
\citep[e.g.,][]{Barnabe2013,Courteau2014}.  Moreover, the IMF may vary with
galaxy type and cosmic time
\citep[]{Vandokkum2008,Conroy2009,Vandokkum2010,Dutton2011a,Conroy2012,Smith2012}.

The dynamics of stars in the inner regions of galaxies (i.e. stellar
kinematics) can be used to infer the total density profiles of
galaxies on scales of about one effective radius. Using integral-field
spectroscopy, the ATLAS3D collaboration obtained resolved stellar
velocity dispersions of nearby elliptical galaxies to constrain
stellar mass-to-light ratios \citep{Cappellari2012}. Another
complementary method is strong lensing, which provides fairly model
independent constraints on the total mass within the Einstein
radius. The combination of stellar kinematics and strong lensing has
proved to be a very powerful approach to constrain the total density
profile of galaxies on scales of about $3$--$9$ kpc
\citep[\eg,][]{Sand2004,Koopmans2006, Gavazzi2007, Jiang2007, Auger2010b, Auger2010a,
Lagattuta2010, Dutton2011b, Newman2013a, Newman2013b, Oguri2014,Sonnenfeld2014}.
Group scale strong lenses (with larger image
separations) have also been used to probe the total density profiles of
galaxies on $\sim10$--$20$ kpc scales
\citep{Kochanek2001, Oguri2006, More2012}. However, strong lensing
systems are rare (at most a handful per square degree) and are
primarily limited to massive early type galaxies at intermediate
redshifts. In this paper, we explore the possibility of using weak
lensing measurements on small scales to potentially overcome these
limitations and to probe the total density profiles of galaxies over a
wide range in redshift and stellar mass \citep[see e.g.,][for limits on stellar
masses from weak lensing around BOSS galaxies]{Miyatake2013,More2014}.

Our goal in this paper is to investigate how well future weak lensing
surveys will be able to measure total density profiles at very small
radial scales. In particular, we will focus on the transition point
where the lensing signal is sensitive to both the stellar component
and the dark matter component of the density profile.  We call this
transition scale the ``equality radius'', noted hereafter as $R_{\rm eq}$. Weak
lensing measurements at the equality radius, in combination with
stellar kinematics on smaller scales, would be a powerful means by
which to constrain both the inner slope of the dark matter density
profile as well as the mass and redshift dependence of the IMF.
To date, there have been few weak lensing measurements at $r<R_{\rm
eq}$. \citet[]{Gavazzi2007} used {\it Hubble Space Telescope} imaging of
22 massive galaxies at $z=0.1$--$0.4$ to measure three data points at
$r<R_{\rm eq}$. The small number of background galaxies at these
separations, however, results in large errors. Here we present
predictions for how these types of measurements will improve as a
function of galaxy mass and redshift for space-based weak lensing
surveys such as {\it Euclid} \citep[]{Laureijs2011} and {\it WFIRST}
\citep[]{Spergel2013}.

The measurement of a weak lensing signal at the equality radius is
inherently difficult. The equality radius is typically about $20$ kpc
(See Section~\ref{subsec:evolution_req}). This leaves only a tiny
radial window in which to find and measure the shapes of background
galaxies. Unbiased measurements of galaxy shapes in crowded
environments poses yet another challenge. The light from the main lens
(or light from neighbouring galaxies that correlate with the lens) may
bias the shape measurements of source galaxies. The investigation of
these biases is a topic of ongoing research, but will not be addressed
in this paper. Instead, here we focus on more simple first-order
questions. We will first investigate how the equality radius varies as
a function of galaxy mass and redshift. Then, we will estimate how
many source galaxies typically lie within this radial scale. We will
also quantify how many source galaxies at $R_{\rm eq}$ have very close
companions and for which shape measurements will be
challenging. Finally, we would also like to understand whether there
is a specific redshift and/or stellar mass range which is most suited
for such measurements.

This paper is organized as follows.  After briefly presenting our
theoretical framework in Section~\ref{sec:theoretical_framework}, we
investigate how $R_{\rm eq}$ varies with stellar mass and redshift and
explore various aspects that determine the signal-to-noise ratio (\sn)
of weak lensing measurements at $R_{\rm eq}$ in
Section~\ref{sec:intuition}. The data analysed in our work is
presented in Section~\ref{sec:data}. Our methodology is presented in
Section~\ref{sec:method}. In section \ref{subsec:proximitycuts} we
explore how cuts based on proximity effects impact the overall source
density -- these estimates are of general interest for all weak
lensing studies including efforts to measure cosmic shear. Finally,
Section~\ref{sec:predict_for_space} shows our predictions for COSMOS
\citep[]{Scoville2007}, {\it WFIRST}, and {\it Euclid}. We summarize our results
and present our conclusions in
Section~\ref{sec:summary_and_conclusions}.

All radial scales are expressed in physical units. The projected
transverse distance from a lens is denoted $r$ whereas $R_{\rm 3D}$
represents a three-dimensional distance. We define dark matter halos
to enclose a spherical overdensity in which the mean density is $200$
times the mean background matter density, $\mtwob=\frac{4}{3}\pi
R_{200{\rm b}}^3 200 \overline{\rho}_{\rm b}^{}\,$, where
$\overline{\rho}_{\rm b}^{}$ is the mean background density and
$R_{200{\rm b}}$ is the halo boundary. For consistency with
\citet[]{Leauthaud2011,Leauthaud2012a} we assume a WMAP5
\citep[]{Hinshaw2009} cosmology with parameters $\Omega_{\rm
  m}=0.258$, $\Omega_{\Lambda}=0.742$, $\Omega_{\rm b}h^2=0.02273$,
$n_{\rm s}=0.963$, $\sigma_{8}=0.796$, $H_{0}=72\,{\rm km\, s^{-1}
  Mpc^{-1}}$.

\section{Theoretical framework}
\label{sec:theoretical_framework}

In this section, we review the necessary theoretical background and
introduce the concept of the ``equality radius'' which will be used
extensively throughout this paper.

\subsection{Density Profiles of Stars and Dark Matter}
\label{subsec:req_def}

Weak gravitational lensing is the deflection of the path of light from
distant galaxies due to the presence of mass over-densities along the
line-of-sight. Weak lensing leads to both a distortion in the shapes
of background galaxies and a magnification of galaxy fluxes and
sizes. These effects are characterized respectively by the shear
$\gamma$ and the convergence $\kappa$. In this work, we focus
specifically on measurements of galaxy-galaxy lensing determined from
the shear $\gamma$.

In the weak lensing regime, the average tangential ellipticity of
background galaxies is related to the tangential component of the shear
$\gamma_{\rm t}$, which in turn is related to the excess surface
density of the intervening mass

\begin{equation}
    \gamma_{\rm t} (r) = \frac{ \overline{\Sigma}(r) - \Sigma(r) }{\Sigma_{{\rm
    crit}}}  = \frac{\Delta \Sigma (r) }{\Sigma_{{\rm crit}}}
    \label{eq:gamma_def}\,,
\end{equation}
where $\overline{\Sigma}(r)$ is the mean projected mass density within
radius $r$ and $\Sigma(r)$ is the projected mass density at radius
$r$. $\Sigma_{{\rm crit}}$ is a critical density defined as:

\begin{equation}
    \Sigma_{{\rm crit}} = \frac{c^2}{4 \pi G} \frac{D_{{\rm os}}}{D_{{\rm ol}}
    D_{{\rm ls}}}\,,
    \label{eq:sigcrit_def}
\end{equation}
where $D_{{\rm os}}$, $D_{{\rm ol}}$ and $D_{{\rm ls}}$ are respectively the angular diameter
distances from the observer to the source, from the observer to the lens, and from
the lens to the source.

The excess surface density $\Delta \Sigma$ is additive and can be
decomposed into three terms representing contributions from dark
matter, gas, and stars within galaxies:

\begin{equation}
    \Delta \Sigma(r) = \Delta \Sigma_{{\rm dm}}(r) + \Delta \Sigma_{{\rm gas}}(r) + \Delta
    \Sigma_{{\rm stellar}}(r) \,.
    \label{eq:delsig_dmbar}
\end{equation}

The gas component, $\Delta \Sigma_{\rm gas}$, may have
contributions from different gas phases such as the cold interstellar
medium (ISM) or the hot X-ray halo gas. As will be discussed further
in Section~\ref{sec:summary_and_conclusions}, for most galaxies in our mass
range, $\Delta \Sigma_{\rm gas}$ is sub-dominant compared to the
stellar and the dark matter contributions. For simplicity, we will
neglect the $\Delta \Sigma_{\rm gas}$ component in this work.

To model the dark matter component, $\Delta \Sigma_{\rm dm}$, we
will assume the standard NFW profile:
\begin{equation}
    \rho(R_{3D}) = \frac{\delta_{c}\overline{\rho}_{\rm b}}{ \left(R_{3D}/R_{\rm s}\right)\left(1+
    R_{3D}/R_{\rm s} \right)^{2}} \,,
    \label{eq:nfw_profile}
\end{equation}
where $R_{\rm s}$ is the characteristic scale radius of the halo. The characteristic density $\delta_{c}$ is given by

\begin{equation}
    \delta_{c} = \frac{200}{3} \frac{c_{\rm 200b}^{3}}{\ln (1+c_{\rm 200b}^{}) -c_{\rm 200b}^{}/(1+c_{\rm 200b}^{})}\,,
\end{equation}
where the concentration parameter $c_{\rm 200b}^{}$ is equal to the ratio
$R_{\rm 200b}/R_{\rm s}$. To
compute $c_{\rm 200b}^{}$ as a function of $\mtwob$ and $z$, we adopt the
concentration mass relation from \citet[]{Maccio2008}.

The $\Delta\Sigma$ profile of an NFW halo is given by
\begin{equation}
    \Delta \Sigma_{{\rm dm}}(r) = r_{{\rm s}} \delta_{c} \overline{\rho}_{{\rm
    b}}^{} g(r/R_{\rm s}) \,, \label{eq:delsig_dmgx}
\end{equation}
where the function $g(x)$ is a dimensionless profile \citep[see][]{Wright2000}.

We model the stellar component using a Hernquist profile
\citep[]{Hernquist1990}. However, over most of the radial range for
which we can measure the shapes of background galaxies, the stellar
component can be approximated by a simple point source term. For
convenience we will adopt this approximation for most of our
calculations. Under this approximation, $\Delta \Sigma_{\rm stellar}$
is simply:

\begin{equation}
    \Delta \Sigma_{{\rm stellar}}(r) \simeq \frac{M_{*}}{\pi r^2} \,.
\label{eq:delsig_stellar}
\end{equation}

\begin{figure*} \centering{
    \includegraphics[scale=2]{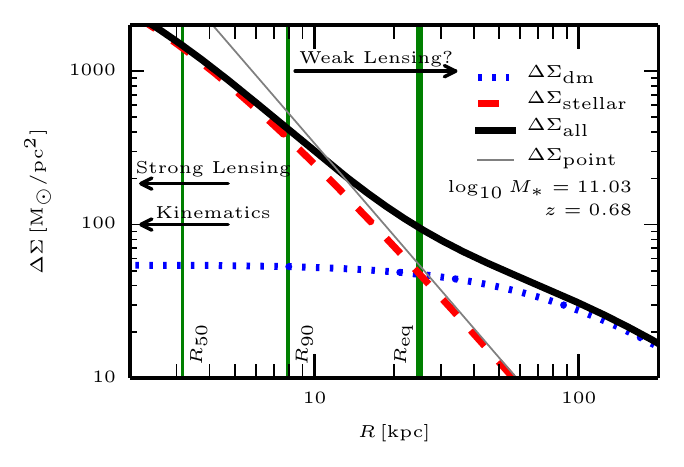}}
    \caption{
    Predicted $\Delta \Sigma$ profile for galaxies with
    $\log_{10}(M^*)\sim 11.3$ and at $z\sim0.68$. The blue dotted line
    represents $\Delta\Sigma_{{\rm dm}}$ assuming an NFW profile. The
    red dashed line represents $\Delta\Sigma_{{\rm stellar}}$ computed
    using a Hernquist mass model. The thin grey line shows
    $\Delta\Sigma_{{\rm stellar}}$ computed using the point source
    approximation (Equation~\ref{eq:delsig_stellar}). The black solid
    line denotes $\Delta\Sigma$ which is the sum of
    $\Delta\Sigma_{{\rm dm}}$ and $\Delta\Sigma_{{\rm stellar}}$.  The
    green thin vertical line labelled ``$R_{90}$'' shows the mean
    radius encompassing $90$ per cent of the flux from lens galaxies.
    Another green thin vertical line labelled ``$R_{50}$'' shows the
    half light radius of lens galaxies. The green thick vertical line
    shows the equality radius $R_{\rm eq}$.  The purpose of this paper
    is to predict the \sn of weak lensing
    measurements on radial scales smaller than $R_{\rm eq}$. }
\label{fig:delsig_profile}
\end{figure*}

Fig.~\ref{fig:delsig_profile} shows $\Delta \Sigma_{\rm stellar}$
computed using both the Hernquist profile and the point source
approximation.

Our goal in this paper is to study the radial scale at which the weak
lensing signal becomes sensitive to the stellar mass of the lens
sample. We define the ``equality radius'' as the radius at which
$\Delta \Sigma_{{\rm dm}}(r) = \Delta \Sigma_{{\rm stellar}}(r)$. This
characteristic radius will be denoted as $R_{\rm eq}$. On scales
smaller than the equality radius, the weak galaxy-galaxy lensing
signal will be dominated by the $\Delta \Sigma_{\rm stellar}$ term and
will hence directly probe $M_{*}$. Fig.~\ref{fig:delsig_profile}
gives an example of a typical $\Delta\Sigma$ profile and the equality
radius $R_{\rm eq}$.

\subsection{Stellar-to-Halo Mass Relation}
\label{subsec:shmr}

To estimate $\mtwob$ from $M_{*}$ for each lens galaxy, we adopt
the stellar-to-halo mass relation (SHMR) from
\citet[]{Leauthaud2012a} (hereafter L12).
In their study, L12 assume a lognormal distribution for the central
galaxy conditional stellar mass function $P(M_*|\mtwob)$.  For
central galaxies that reside in halos of mass $\mtwob$, the average
of their logarithmic stellar mass, $f_{\rm SHMR}(\mtwob)$, satisfies the following equation:

\begin{equation}
\begin{aligned}
    &\log_{10} \left(f_{{\rm SHMR}}^{-1} (M_{*})\right) = \log_{10}
    \left(\mtwob\right) \\
    &= \log_{10}\left(M_{1}\right) + \beta \log_{10} \left(\frac{M_{*}}{M_{*,0}}\right)
    +
    \frac{\left(M_{*}/M_{*,0}\right)^{\delta}}{1+\left(M_{*}/M_{*,0}\right)^{-\gamma}}
    - \frac{1}{2}\,.
\end{aligned}
\end{equation}
This functional form with parameters $M_1$, $M_{*,0}$, $\beta$,
$\delta$, and $\gamma$ is motivated from \citet[]{Behroozi2010}. We
refer the reader to \citet[]{Leauthaud2011} and \citet[]{Behroozi2010}
for the definitions of these parameters, notations, and how these
parameters control the shape of SHMR.

L12 fit this model to the abundances of galaxies, their clustering, and
the weak lensing signal from the COSMOS field in three redshift bins
($z\in[0.22,0.48]$, $[0.48,0.74]$, $[0.74,1.00]$). Their results
suggest a very weak evolution of the global SHMR from $z=0.2$ to $z=1$,
especially for low stellar mass galaxies with $M_{*} \lesssim 2 \times
10^{10}$ \msun. We use the SHMR parameters from L12 in each redshift
bin.

Note that this SHMR is only valid for central galaxies. According to
the modelling results from L12, about 20 per cent of galaxies in the
COSMOS field
are expected to be satellites.  For our predictions, however, we
neglect satellite galaxies and make the simplifying assumption that
all lens galaxies are centrals.

The average halo mass of galaxies of a given stellar mass
can be computed using $P(\mtwob|M_*)$, which is related to
$P(M_*|\mtwob)$ via Bayes' theorem
\begin{eqnarray}
    P(\mtwob|M_*)&=&\frac{P(M_*|\mtwob)P(\mtwob)}{P(M_*)} \nonumber\\
                            &\propto &P(M_*|\mtwob)n(\mtwob)\,,
    \label{eq:pof_mh_givenmstar}
\end{eqnarray}
where $n(\mtwob)$ represents the halo mass function. For this
analysis, we will use the \citet[]{Tinker2008} halo mass function. We
calculate the average halo mass of galaxies within a given stellar
mass bin $[M_1,M_2]$ as:
\begin{equation}
    \langle \mtwob\rangle_{[M_1,M_2]}=
\frac{\int_{M_1}^{M_2} dM_*
\int_{0}^{\infty} \mtwob P(\mtwob|M_*) d\mtwob }{
	\int_{M_1}^{M_2} dM_*
\int_{0}^{\infty} P(\mtwob|M_*) d\mtwob} \,.
    \label{eq:mh_givenmstar}
\end{equation}
The variation of $\mtwob$ as a function of lens stellar mass and redshift
is shown in Fig.~\ref{fig:various_effects}.

\section{Small scale Weak lensing: Main effects}
\label{sec:intuition}

Our goal is to predict the expected \sn of weak lensing measurements
within the equality radius $R_{\rm eq}$. A variety of different effects
will impact the expected weak lensing signal
at this radial scale. We list these effects below.

\begin{enumerate}
\item \req depends on the SHMR, the concentration-mass relation, and
  their evolution with redshift.
\item The angular diameter distance and its dependence on redshift
  determines the apparent area on the sky covered by a given $R_{\rm
    eq}$. For a fixed number of lens galaxies, a larger apparent area
  on the sky will correspond to a larger number of source galaxies and
  hence a higher \sn.
\item For a fixed survey area, the number of lens galaxies will increase
  with redshift because the survey covers a larger comoving volume.
\item The strength of the lensing signal (i.e, the shear) depends on
  the value of $\Sigma_{\rm crit}$ which depends on the redshift of
  the lens and the source. For a fixed source redshift, $z_{\rm s}$, the
  strength of the lensing signal peaks around $z_{\rm s}/2$.
\item The strength of the shear around lens galaxies will be large at
  $R_{\rm eq}$. We need to be mindful of the regime where lensing is
  no longer weak.
\item Magnification bias may alter the observed source density.
\item The amount of ``real estate''  between the radius at which the
  light from a lens becomes insignificant and \req depends on the
  sizes of lens galaxies.
\item Unbiased measurements of shapes are difficult for galaxies that
  are blended or that are close to other galaxies so proximity effects
  are important.
\end{enumerate}

Some of these trends ((i)-(vi)) can be computed analytically. However
(vii) and (viii) are non trivial. In particular, quantifying the
effects of close pairs is especially difficult. For this reason, our
predictions will be largely based on numbers drawn directly from a
COSMOS weak lensing catalogue (see Section~\ref{sec:data}). Below, we
discuss each of these effects in further detail with an emphasis on
developing an intuition for how each effect drives our predicted
\sn. We also clarify which components of our model are computed
analytically and which components are drawn from real data.

\subsection{Equality radius}
\label{subsec:evolution_req}
As defined in Section~\ref{subsec:req_def}, $R_{\rm eq}$ corresponds
to the radius where the contributions of the dark matter component and
the stellar component to $\Delta \Sigma$ are equal. The \sn of weak 
lensing measurement on such small scales
depends on the number of background source galaxies that lie within a
projected distance $r<R_{\rm eq}$. Hence, a larger value of $R_{\rm
  eq}$ should result in a larger \sn.

\begin{figure*}
    \minipage[t]{0.31\textwidth}
    \includegraphics[scale=0.8]{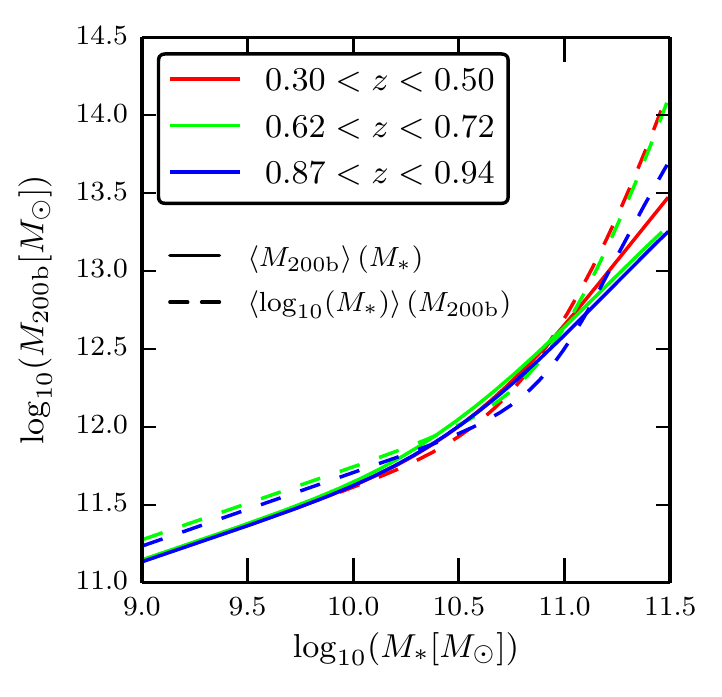}
    \endminipage%
    \minipage[t]{0.31\textwidth}
        \includegraphics[scale=0.8]{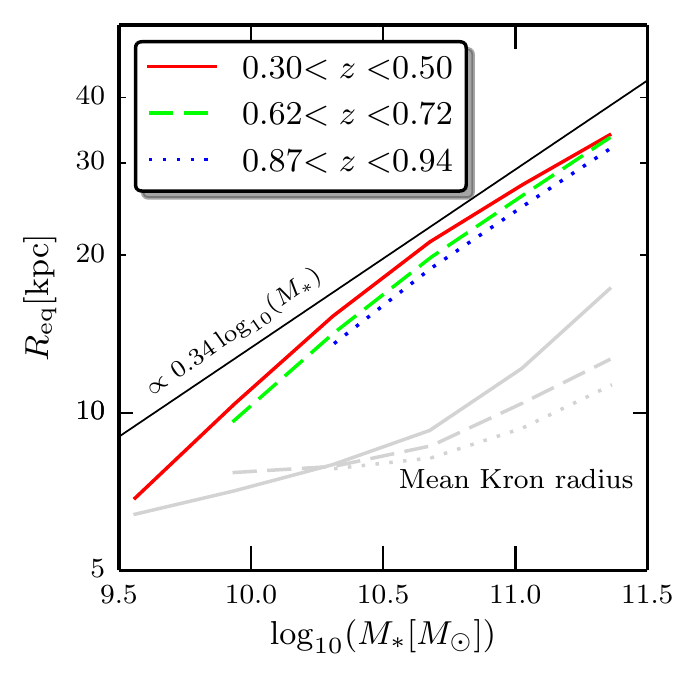}
    \endminipage%
    \minipage[t]{0.31\textwidth}
        \includegraphics[scale=0.8]{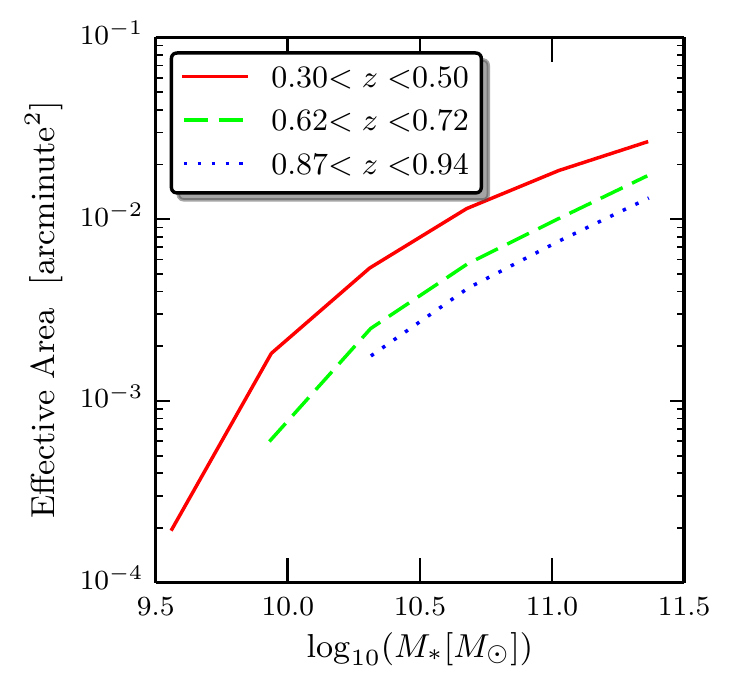}
    \endminipage%
    \caption{Left: variation of $\mtwob$ as a function of stellar mass
      and redshift based on SHMR. Coloured lines represent three
      different redshift bins for lens galaxies.  Solid lines show
      $\left<\mtwob\right>$ whereas dashed line show
      $\left<\log_{10}(M_{*})\right>$. Differences between the solid
      and dashed lines are due to scatter in the SHMR
      relation. The green solid line
      corresponding to the redshift range ($0.62<z<0.72$) has been shifted
      horizontally by $0.03$ dex for presentation purposes. Middle:
      evolution of $R_{\rm eq}$ as a function of stellar mass and redshift.
      Coloured lines represent three
      different redshift bins for lens galaxies. Grey lines indicate
      the geometric mean of the semi-major and the semi-minor length
      of the Kron ellipses of lens galaxies.  As a reference, the
      typical half light radius of lens galaxies at $z\sim0.4$ is
      $R_e=2.4$ kpc for $M_{*} \approx 5.5 \times 10^{9} \msun$ and
      $R_e=3.5$ kpc for $M_{*} \approx 1.3 \times 10^{11}
      \msun$. Right: evolution of the effective area within $R_{\rm
        eq}$ as a function of stellar mass and redshift. }
    \label{fig:various_effects}
\end{figure*}

We compute $R_{\rm eq}$ by equating the right hand sides of
Equations~\ref{eq:delsig_dmgx} and \ref{eq:delsig_stellar}.
Fig.~\ref{fig:various_effects} shows the variation of $R_{\rm
  eq}$ as a function of lens stellar mass and redshift.  At a given
redshift, $R_{\rm eq}$ increases with lens stellar mass. The increase
in \req with stellar mass is in contrast to the $M_*/\mtwob-\mtwob$
relation, which shows a distinct peak at a preferred halo mass scale
of $M_{\rm 200b} = 10^{11.8}\msun$ (e.g., \citealt[]{Behroozi2010}; L12;
\citealt[]{RodriguezPuebla2013}).

This trend can be easily understood in the following manner. In the
inner regions of halos ($r\lesssim r_{{\rm s}}$), the profile function
in Equation~\ref{eq:delsig_dmgx}, $g(r/r_{\rm s})\approx
1$. Therefore, we obtain:
\begin{equation}
    R_{\rm eq} \propto \left( \frac{M_{*}}{\ave{\mtwob}^{\frac{1}{3}}} \right)^{\frac{1}{2}} \overline{\rho}_{\rm b}^{-\frac{1}{3}}
    \,.
    \label{eq:req_propto}
\end{equation}
For a fixed redshift, $R_{\rm eq}$ is proportional to the first term in the
above equation.  We compute $\ave{\mtwob}$ from
Equation~\ref{eq:mh_givenmstar}, and find that the power law slope of the
$\ave{\mtwob}-M_{*}$ relation is approximately unity at a stellar mass
of about $\log_{10}(M_{*})=10.5$,
which implies the approximate scaling $\log_{10}(R_{\rm
eq}) \propto 0.34 \times \log_{10}(M_{*})$. This scaling is shown
in Fig.~\ref{fig:various_effects} by a solid black line.
It agrees with the increase of \req with
stellar mass that we observe.

\subsection{Effective Area within $R_{\rm eq}$}
\label{subsec:effective_area}

We will use the Kron ellipse (which roughly encompasses 90 per cent of the
light) to define the spatial extent of lens galaxies (see
Section~\ref{subsec:proximitycuts}). We define the effective area,
$A_{\rm eff}$, as the angular area between the outskirts of
lens galaxies (as traced by the Kron ellipse) and $R_{{\rm eq}}$ --
this is a measure of how much ``real estate'' is available for weak
lensing measurements at $R_{\rm eq}$. Fig.~\ref{fig:various_effects}
shows how $A_{\rm eff}$ varies as a function of lens stellar mass and
redshift. The redshift dependence of $A_{\rm eff}$ is driven by the
angular diameter distance. A fixed value of $R_{{\rm eq}}$ corresponds
to a larger effective area at lower
redshifts. Fig.~\ref{fig:various_effects} shows that massive
galaxies at low redshift have the largest effective area and by
consequence, the largest number of source galaxies per lens within the
equality radius.

\subsection{Geometric Effects due to $\Sigma_{\rm crit}$}
\label{subsec:geometric}

The strength of the lensing signal depends on the value of
$\Sigma_{\rm crit}$, which depends on the redshifts of both lens and
source galaxies. To gain an intuition for how $\Sigma_{\rm crit}$
drives \sn, let us consider a simple scenario in which we have a
{\it fixed} number of lens galaxies $N_{\rm l}$ and a fixed high
redshift source plane $z_{\rm s}$. The projected number density of sources
is denoted as $n_{\rm s}$. The error on $\Delta \Sigma$ within a fixed
physical aperture $R_{\rm ap}$ scales as $\sigma_{_{\Delta \Sigma}}
\propto \Sigma_{{\rm crit}} \, N_{{\rm pairs}}^{-1/2}$ where $N_{\rm
  pairs}$ is the number of lens-source pairs. The number of
lens-source pairs is simply:

\begin{equation}
    N_{\rm pairs} = N_{\rm l} \times n_{\rm s} \times \pi R_{\rm ap}^{2} /
    D_{\rm ol}^{2} \,.
    \label{eq:aperture_pairs}
\end{equation}
Putting these two together, the $D_{{\rm ol}}$ term cancels and the
error on $\Delta \Sigma$ within a fixed physical aperture $R_{\rm ap}$
scales as:

\begin{equation}
    \sigma_{_{\Delta \Sigma}} \propto \frac{D_{{\rm os}}}{D_{{\rm ls}}}. \label{eq:errdelsig_scale}
\end{equation}
For a fixed source redshift, a lower lens redshift results in a larger
$D_{{\rm ls}}$, which corresponds to smaller error on $\Delta
\Sigma$. Therefore, from a pure geometric point of view, the \sn for
weak lensing measurements within a fixed aperture will {\it increase}
at lower redshifts. This may seem in contrast to the common intuition
that the \sn of weak lensing measurements peaks at about a redshift
of $z_{\rm s}/2$. The reason for this difference is simply that Equation
\ref{eq:errdelsig_scale} assumes a fixed number of lens galaxies
$N_{\rm l}$. In a weak lensing {\it survey} however, the number of
lens galaxies per unit comoving line-of-sight distance will decrease
towards low redshifts (causing $N_{\rm pairs}$ to also decrease).

In this paper, we are primarily interested in predictions for the
\sn of weak lensing measurements for future weak lensing
surveys. However, we note that Equation \ref{eq:errdelsig_scale}
implies that another strategy for maximizing the \sn of small scale
weak lensing measurements is targeted observations of very low
redshift massive galaxies (\citealt[]{Gavazzi2007}; Okabe et al. in
prep.)

\subsection{Amplitude of Shear on Small Scales}
\label{subsec:weakshear_regime}

As we try to push towards small radial scales, the weak shear
assumption, $|\gamma|<<1$, may no longer be valid. Most of the shear
measurement pipelines are designed with cosmic shear studies in mind
and are not necessarily well tested for large shear values.  For
example, the GRavitational lEnsing Accuracy Testing 3 (GREAT3)
competition \citep[]{Mandelbaum2014} only tested $\gamma$ up to values
of $0.05$. In this section we calculate the amplitude of the shear
signal at the equality radius to determine whether or not this still
corresponds to a weak lensing regime.

\begin{figure*}
    \minipage[t]{0.39\textwidth}
        \includegraphics{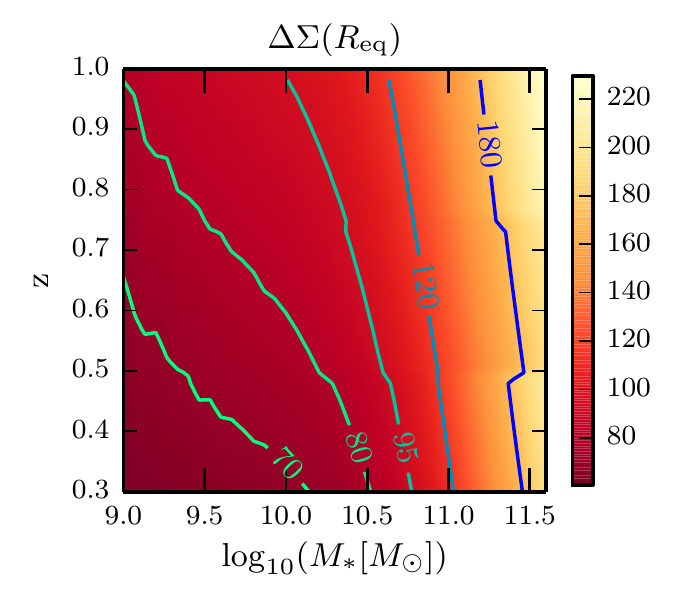}
    \endminipage%
    \minipage[t]{0.39\textwidth}
        \includegraphics{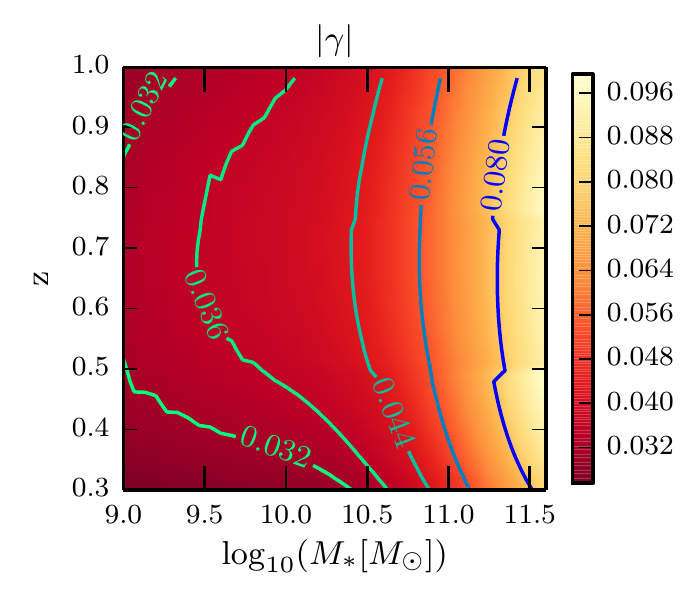}
    \endminipage%
    \caption{Left: variation of $\Delta \Sigma$ at $R_{\rm eq}$ as a
      function of lens stellar mass and redshift assuming source
      galaxies at $z_{\rm s}=2$. $\Delta\Sigma(R_{\rm eq})$ monotonically
      increases with stellar mass and redshift. Right: same as the
      left panel but for the shear $\gamma$.  At fixed stellar mass,
      $\gamma$ reaches maximum values at $z\sim0.6$.}
    \label{fig:signal_dist}
\end{figure*}

At $r=R_{\rm eq}$, $\Delta \Sigma$ is simply:
\begin{equation}
\Delta \Sigma(R_{\rm eq})=2\times\frac{M_*}{\pi R_{\rm eq}^2} \,.
\end{equation}
The left hand panel of Fig.~\ref{fig:signal_dist} shows the
dependence of $\Delta \Sigma(R_{\rm eq})$ on lens stellar mass and
redshift. As discussed in Section~\ref{subsec:evolution_req}, at fixed
redshift, $R_{\rm eq}$ increases with stellar mass but with a shallow
slope of $0.34$. Therefore, $\Delta \Sigma(R_{\rm eq})$ increases with
stellar mass with a power law slope of $0.32$. Similarly, at a fixed
stellar mass, \req decreases with redshift, leading to an increase in
$\Delta \Sigma(R_{\rm eq})$ with redshift.

We now estimate the typical value of $\gamma$ at $r=R_{\rm eq}$. For
simplicity, we assume that all source galaxies are located at a fixed
redshift $z_{\rm s}=2$. The results are shown in the right hand panel of
Fig.~\ref{fig:signal_dist}. The difference between the contour shapes
in the left and the right panel are due to the dependence of
$\Sigma_{\rm crit}$ on the lens redshift; for fixed $z_{\rm s}=2$,
$\Sigma_{\rm crit}$ reaches a minimum at intermediate redshifts. We
find that galaxies with $M_*>10^{11} \msun$, on average, will have
shear values of order $0.05$--$0.01$ at $R_{\rm eq}$. Hence, these types
of measurements will require a careful calibration of shear
measurements up to values of about $0.1$. Biases in the shear
measurement at such high values are currently being investigated by
the ARCLETS collaboration\footnote{{\sc
    ARCLETS:}\url{http://www.het.brown.edu/people/ian/ClustersChallenge/}}.

Moreover, in this shear regime, the shapes of source galaxies
  are described by a combination of the shear and the convergence,
  called the reduced shear \citep[]{Schneider1995}. The
  interpretation of the small-scale weak lensing will require modelling of the
  reduced shear rather than the shear (that we investigate here).
  However, the difference between the two is expected to be relatively minor ($5$--$10$ per
  cent) on the radial scales and the halo mass scales considered in our work
  \citep[\eg,][]{Mandelbaum2006,Johnston2007,Leauthaud2010}.

An important point to note here is that for our calculations, we are
considering the value of the shear for the {\it average} halo at fixed
lens stellar mass.  The type of measurements that we are proposing
here are not possible at the hearts of rich groups or clusters of
galaxies at intermediate redshifts where the weak lensing
approximation is clearly not valid. For massive galaxies, however,
there is a large scatter in $\mtwob$ at fixed stellar mass (see \eg,
\citealt[][]{More2011}; L12). For example, at a fixed stellar mass of
$\log_{10}(M_{*})=11$, there is a $0.46$ dex spread (a factor of $2.9$) in halo
mass, and a $0.7$ dex spread (a factor of $5.0$) at $\log_{10}(M_{*})=10.5$.
Our calculations are valid for the average halo at fixed stellar mass -- not for
galaxies in rich groups or clusters.

\subsection{Proximity Effects and Blends}
\label{subsec:proximity}

To measure weak lensing signals, accurate and unbiased measurements of
both shapes and photometric redshifts are critical. Such measurements
are difficult when the angular separation between the lens and the
source galaxy is small. In particular, the equality radius is merely a
factor of 2-3 larger than the typical size of lens galaxies (as traced
by the Kron ellipse, see Fig.~\ref{fig:various_effects}). If a source galaxy is located too
close to a lens, then the light from the lens galaxy may contaminate
the photometry as well as the shape measurement of the source
galaxy. Biases in shape measurements will directly translate into
biases in the measured weak lensing signal.  Biases in the photometry
contribute indirectly via biases in photometric redshifts which are
required to compute $\Sigma_{\rm crit}$. The magnitude of these
proximity effects depend on the size distribution of lens galaxies but
also on the size distribution of neighbouring galaxies that correlate
with the lens sample. We will use existing COSMOS ACS data to
estimate the magnitude of these effects. We will devote large parts of
Section~\ref{sec:method} in order to quantify the magnitude of
proximity effects.


\section{Data}
\label{sec:data}
As discussed in the previous section, among the various factors
affecting the expected \sn of weak lensing
measurements on small scales, the number of source galaxies lost due
to proximity effects are the hardest to quantify analytically.  It is
primarily for this reason that we opt to use a real catalogue for the
basis of our study instead of taking a semi-analytical approach such
as the one explored by \citet{Dawson2014}.
We will use the COSMOS ACS catalogue (described in the subsequent
section) as our primary catalogue with which to conduct this analysis.
The COSMOS field is small which means that our choice represents a
trade-off between area and high resolution imaging but, as we will
show further on in this paper, proximity effects are the largest
determining factor for our predictions which motivates our choice. We
now describe the data products that we use in greater detail.

\subsection{COSMOS Weak Lensing Catalogue}
\label{subsec:weaklensingcatalog}
The COSMOS program has imaged the largest contiguous area (1.64
degrees$^2$) with the {\it Hubble Space Telescope (HST)} using the
{\it Advanced Camera for Surveys (ACS)} {\it Wide Field Channel (WFC)}
\citep[][]{Scoville2007}.  The $5\sigma$ point-source limiting depth
is $I_{\rm AB} = 27.2 \ \mathrm{mag}$ in a 0\farcs24 diameter aperture
\citep[][]{Koekemoer2007} and
the size of point spread function (hereafter PSF) is $0.12''$ \citep[][]{Leauthaud2007}.
This combination of depth and exquisite resolution over a moderately wide
area motivates our choice to use this as our primary data set. The
details of the COSMOS weak lensing catalogue have already been described
in ample detail elsewhere (\citealt[]{Massey2007a, Leauthaud2007,
  Rhodes2007}; L12). The COSMOS weak lensing catalogue contains $3.9
\times 10^5$ galaxies with accurate shape measurements, which
represents a number density of $66$ source galaxies per
arcminute$^{2}$.

For shape measurements, the COSMOS weak lensing catalogue uses the RRG
algorithm \citep[]{Rhodes2000}. The exact details of the shape
measurement procedure are mostly un-important for this paper. We do
however use information about {\em when} a shape measurement was
possible with the expectation that the failure rate of galaxy shape
measurement will increase for close lens-source pair
configurations. The RRG COSMOS catalogue does not explicitly flag source
galaxies that have close companions. However, galaxies may fail to
converge on a shape measurement if the RRG algorithm fails to determine a
centroid (which typically occurs for blends). This is in contrast to
the {\sc lensfit} algorithm, for example, which explicitly rejects galaxies
in close pairs whose isophotes are overlap at two-sigma level of the
pixel noise \citep[]{Miller2007, Miller2013}.

To select galaxies with precise shape measurements, we adopt the same
four selection cuts as \citet[]{Leauthaud2007}. In addition, we
restrict ourselves to galaxies that are detected in the COSMOS Subaru
catalogue (See Section~\ref{subsec:photoz}) and with $0<\sigma_{\rm
  meas}<0.2$, where $\sigma_{\rm meas}$ represents the shape
measurement error from \citet[]{Leauthaud2007}. We also remove a small
fraction of galaxies for which SourceExtractor \citep[]{Bertin1996}
failed to measure the ${\rm KRON\_RADIUS}$ (see
Section~\ref{subsec:proximitycuts} for the definition of this
parameter). These additional cuts remove $29$ per cent of galaxies from the
COSMOS weak lensing catalogue, leaving $2.7 \times 10^5$ galaxies ($45$
galaxies per arcminute$^2$) with shape measurements and which are
detected by the COSMOS Subaru observations.

\subsection{COSMOS Photo-z Catalogue}
\label{subsec:photozcatalog}

Photometric redshifts (hereafter, ``photo-zs'') are necessary in order
to separate foreground and background galaxies. Ideally, our photo-zs
would be determined from multi-band imaging matched in both resolution
and depth to the COSMOS F814W imaging. However, the COSMOS photo-zs
are measured from ground-based imaging with a PSF that is larger by
about a factor of $10$ compared to the ACS data. In
Section~\ref{sec:method}, we study how this photo-z matching affects
our predictions.

For this paper, we use the COSMOS photo-z catalogue version 1.8 presented
in \citet{Ilbert2009}. These photo-zs have been derived using
a $\chi^2$ template fitting method using over 30 bands of multi-wavelength
data from UV, visible near-IR to mid-IR, and also have been calibrated
with large spectroscopic samples from VLT-VIMOS \citep[]{Lilly2007}
and Keck-DEIMOS.

\subsection{Stellar Mass Estimates}
\label{subsec:stellarmass}

Stellar mass estimates are required in order to predict
$\Delta\Sigma$. We use the same stellar mass estimates as L12 and
refer the reader to L12 and \citet{Bundy2010} for further details.

Stellar mass estimates are based on PSF-matched 3\farcs0 diameter
aperture photometry from the ground-based COSMOS catalogue (filters
$u^*, B_J, V_J, g^+, r^+, i^+, z^+, K_s$)
\citep[][]{Capak2007,Ilbert2009, McCracken2010}.  The depth in all
bands reaches at least 25th magnitude (AB) except for the $K_s$-band,
which is limited to $K_s < 24 \, [\mathrm{mag}]$.  Stellar masses are
derived using the Bayesian code described in \citet{Bundy2006},
which uses \citet[]{Bruzual2003} population synthesis code and
assume a Chabrier IMF \citep[]{Chabrier2003} and a
\citet[][]{Charlot2000} dust model.  The stellar mass completeness is
defined from magnitude limits $K_s < 24 \, [\mathrm{mag}]$ and $I_{{\rm
    814W}} < 25 \,[\mathrm{mag}]$.  For the redshift range that we are
interested in here ($0.2 \leq z \leq 1.0$), the stellar mass
completeness ranges from $M_{*} = 10^{8.16}$\msun to $10^{9.98}$\msun.


\section{Method}
\label{sec:method}
Here we describe the method that we use to compute the
predicted \sn for $\Delta\Sigma (r<R_{\rm eq})$ for
weak lensing data sets with COSMOS like quality. For a given set of
foreground lens galaxies, the main ingredients necessary to make this
prediction are the number of lens galaxies, the number of background
galaxies with shape measurements as a function of distance, the shape
noise for these background galaxies, and the redshift distribution of
source galaxies. We describe how we extract these quantities from the
COSMOS ACS catalogue.

\subsection{Lens and Source Samples}
\label{subsec:lens_and_source_sample}
There is some arbitrariness involved in the definition of the
foreground lens sample used to perform a \sn investigation. 
The predicted value of the \sn
depends on the number of galaxies in the lens sample, which depends on
the choice of the binning scheme. For example, equal-sized redshift
bins will provide a larger volume for higher redshift bins, which
results in a larger number of galaxies in the higher redshift bins. In
our work, we opt to divide our galaxies into $7$ equal comoving volume
bins in redshift and $6$ logarithmically spaced stellar mass bins. In
each redshift bin, we only consider bins that are complete in stellar
mass (Section~\ref{subsec:stellarmass}; L12). The characteristics of
our lens samples are provided in Table~\ref{table_nl}.

\begin{table*}
    \begin{tabular}{c|c|c|c|c|c|c|}
       \hline
       \hline
        Redshift & \multicolumn{6}{|c|}{$\log_{10} M_{*} [{\rm M_{\odot}}]$}\\
& $[$ 9.37 , 9.74 $]$& $[$ 9.74 , 10.11 $]$& $[$ 10.11 , 10.49 $]$& $[$ 10.49 , 10.86 $]$& $[$ 10.86 , 11.23 $]$& $[$ 11.23 , 11.60 $]$\\
\hline 
 \hline
$[$ 0.30 , 0.50 $]$ & 2477 & 2000 & 1690 & 1404 & 731 & 162 \\ 
$[$ 0.50 , 0.62 $]$ & 1688 & 1185 & 1110 & 878 & 494 & 113 \\ 
$[$ 0.62 , 0.72 $]$  & $\times$ & 1476 & 1209 & 1144 & 750 & 212 \\ 
$[$ 0.72 , 0.80 $]$  & $\times$ & 1669 & 888 & 767 & 479 & 137 \\ 
$[$ 0.80 , 0.87 $]$  & $\times$  & $\times$ & 1436 & 1458 & 943 & 266 \\ 
$[$ 0.87 , 0.94 $]$  & $\times$  & $\times$ & 1107 & 993 & 622 & 173 \\ 
$[$ 0.94 , 1.00 $]$  & $\times$  & $\times$ & 1185 & 1058 & 737 & 241 \\ 
 \hline 
Mean number & 2007.8 & 1526.4 & 1188.6 & 1049.9 & 645.9 & 171.9 \\

        \hline
        \hline
    \end{tabular}
    \caption{Number of lens galaxies from our COSMOS catalogue in bins
      of stellar mass and redshift. Bins that are incomplete in terms
      of stellar mass are shown as $\times$.
      The numbers in the bottom row are the mean values after the
      correction of sample variance described in
      Section~\ref{subsec:lens_and_source_sample}.}
    \label{table_nl}
\end{table*}

Despite our choice of redshift bins with equal comoving volumes, the
number of lens galaxies within each stellar mass bin fluctuates due to
sample variance.  We would like to minimize the effects of such
variations on the calculated \sn. We make the
assumption that the total stellar mass function of galaxies does not
strongly evolve at $z<1$ (motivated by \citealt[]{Bundy2010,
  Ilbert2010}; L12; \citealt[]{Moustakas2013}). We fit a constant to
the number of galaxies in each stellar mass bin, assuming Poisson
statistics. The value of these constants are provided in
Table~\ref{table_nl} as ``mean number''. 
We use this mean number in order to
generate our predictions.

As our input catalogue for source galaxies, we use the COSMOS ACS weak
lensing catalogue.  As described in
Section~\ref{subsec:weaklensingcatalog}, we restrict ourselves to
source galaxies that pass a combination of three quality cuts as well
as four quality cuts described in \citet[]{Leauthaud2007}.  These
galaxies form our source catalogue.

\subsection{Impact of Proximity Effects on the Overall Source Density}
\label{subsec:proximitycuts}

In this section we explore how cuts based on proximity effects impact
the overall source density. These estimates are of general interest
for all weak lensing studies including efforts to
measure cosmic shear for example.

Galaxies in close pairs may have biased shear estimates, but the exact
level of such bias or how it varies as a function of the distance
between close pairs, remains poorly understood. As a result, most
current weak lensing pipelines do not necessarily have a well
justified criterion to determine which galaxies have shapes that may
be biased because of neighbouring galaxies. For example, the ACS
lensing catalogue that we use in this paper makes no stringent cuts to
remove source galaxies in close-pair configurations, even though these
galaxies may have biased shear estimates. On the other hand,
\citet[]{Miller2013} take a more conservative approach while
constructing the CFHTLS weak lensing catalogue (see
Section~\ref{subsec:weaklensingcatalog}).  However, both of these
choices remain subjective and do not study the amount of shear bias in
close pairs.  Quantitative investigation of shear bias is a topic of
active research, but is beyond the scope of this paper. Instead, we
use a simple and easily reproducible criterion for identifying close
pairs and we explore how the predicted \sn for
small-scale lensing measurements varies for source galaxy selections
that are more or less conservative.

Our scheme to identify close pairs of galaxies is based on the Kron
parameters from SourceExtractor \citep[]{Bertin1996} (hereafter
\sext), which generalizes the luminosity-weighted radius of
\citet{Kron1980} to elliptical apertures.  Our choice is motivated by
the fact that Kron parameters are often available in imaging
catalogue, which makes our criterion easily reproducible. We note,
however, that our procedure will not identify blends at a fixed
isophotal level. While an isophotal definition of blends would have been
preferable, it was not possible to calculate such a criterion given
the catalogues available to us.

The Kron parameters consist of ${\rm A\_IMAGE}$, ${\rm B\_IMAGE}$, ${\rm KRON\_RADIUS}$, and ${\rm THETA\_WORLD}$. These parameters can
be used to define a ``Kron ellipse'' for each galaxy that roughly
encompasses $90$--$95$ per cent of the flux. For \sext, 
${\rm KRON\_RADIUS}$ is computed by multiplying the first moment radius
by a factor of $2.5$, then the major
axis and the minor axis of the Kron ellipse are given by ${\rm
  A\_IMAGE \times KRON\_RADIUS}$ and ${\rm B\_IMAGE \times
  KRON\_RADIUS}$ respectively. 
The position angle of the Kron
ellipse, ${\rm THETA\_WORLD}$, is measured between the right ascension
and the direction of the semi-major axis.

\begin{figure} \centering{
    \includegraphics[width=\columnwidth,keepaspectratio]{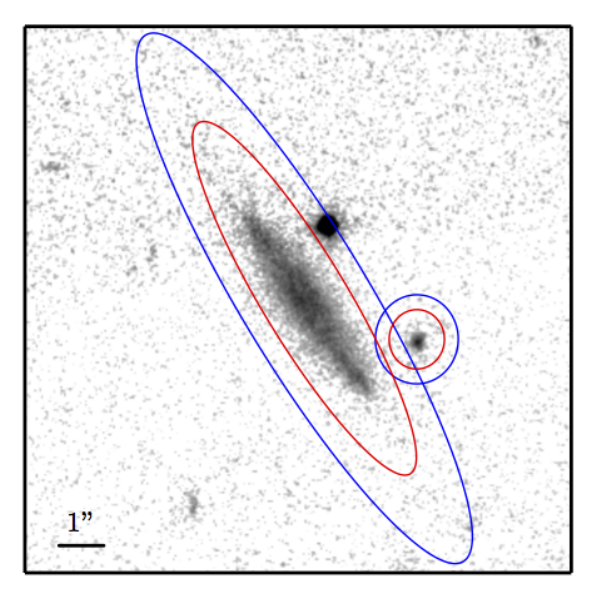}}
  \caption{Two neighbouring galaxies with Kron ellipses overlaid. For
    the blue ellipses, the Kron ellipse has been scaled by a factor of
    $f=1.2$. For the red ellipses, the Kron ellipse has been scaled by
    a factor of $f=0.8$. These galaxies are rejected from our source
    catalogue when $f=1.2$ but are not rejected when $f=0.8$. The bright
    compact object is a star and does not belong in the source
    catalogue.}
\label{fig:kronoverlap_example}
\end{figure}

We place Kron ellipses around all galaxies and identify galaxies with
overlapping Kron ellipses. Our baseline predictions reject all source
galaxies with overlapping Kron ellipses. In order to make this
rejection criterion more or less conservative, we simply increase or
decrease the size of the ellipse by multiplying the major and minor
axis by a constant factor $f$. This procedure is illustrated in
Fig.~\ref{fig:kronoverlap_example}.  The subsample of galaxies
rejected by our criterion, when the multiplicative factor is $f$, will
be denoted by $\sc{R}_f$. Those galaxies which are rejected when
$f=f_2$ but not rejected when $f=f_1<f_2$, will be denoted by
$\sc{R}_{[f_1,f_2]}$. We use $f=1.0$ for our fiducial set of
predictions and we explore how our predictions vary for different
values of $f$. Table~\ref{table_ngal} shows the number and fraction of
source galaxies that are rejected for various values of $f$.

Table~\ref{table_ngal} shows that the overall source density is very
sensitive to proximity cuts. Rejecting all source galaxies that have a
Kron ellipse that overlaps with a neighbouring galaxy leads to a $20$ per cent
decrease in the overall source density. Adopting more conservative
values of $f=1.2$ will lead to a $30$ per cent decrease in the overall source
density. Understanding how neighbouring galaxies impact shear bias and
which source galaxies need to be rejected is clearly of importance for
all weak lensing studies, not just the particular science application
discussed in this paper.

\begin{table}
    \begin{center}
    \begin{tabular}{c|c|c|c|c|c|}
       \hline
       \hline
          & $\sc{R}_{[0, 0.5]}$ & $\sc{R}_{[0, 0.8]}$ & $\sc{R}_{[0, 1.0]}$ & $\sc{R}_{[0, 1.2]}$ & $\sc{R}_{[0, 1.5]}$\\
\hline 
 \hline 
 Number & 16993 & 43182 & 61023 & 80473 & 111577 \\
Fraction(\%) & 6.3 & 16.0 & 22.6 & 29.8 & 41.3 \\

        \hline
        \hline
    \end{tabular}
    \caption{The number and fraction of
    galaxies identified as close pairs in the source catalogue for
    different values of $f$. Our default scheme is $f=1$.
    In this case, $22.6$ per cent of galaxies are rejected from the overall source catalogue.}
    \label{table_ngal}
\end{center}
\end{table}

\begin{figure*} \centering{
\includegraphics[scale=1.3]{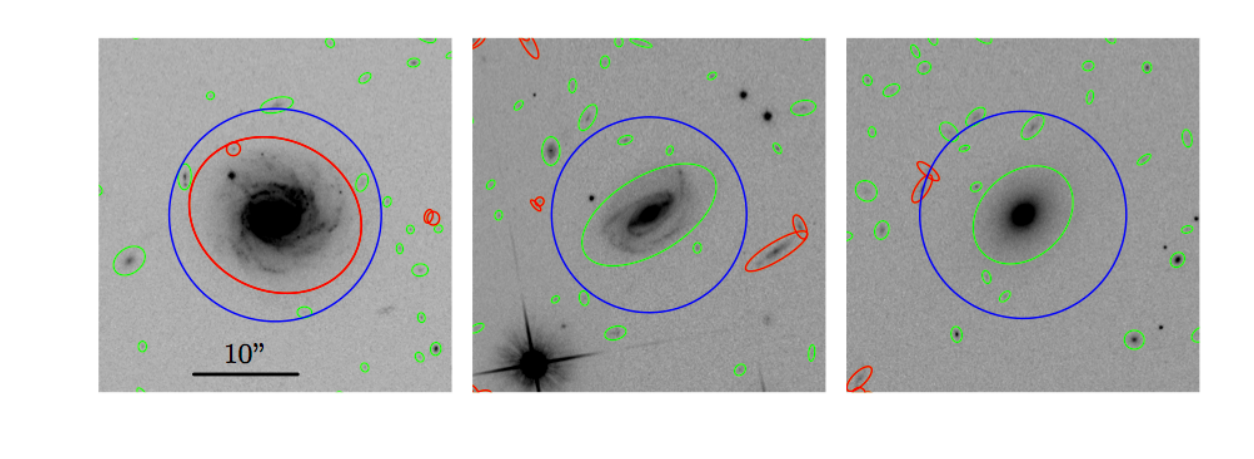}}
\caption{The three large galaxies at the centre of each postage stamp
  are examples of lens galaxies from the COSMOS survey at $z\sim0.25$
  and with $\log_{10} M_{*}[{\rm M_{\odot}}]\sim 11.1$. Green ellipses
  show Kron ellipses (here we use $f=1$). Blue circles show $R_{\rm
    eq}$.  Galaxies with overlapping Kron ellipses are identified by
  red ellipses -- these are rejected from our source catalogue. No
  circles or ellipses are drawn for stars.}
\label{fig:image_hst}
\end{figure*}

Fig.~\ref{fig:image_hst} shows an example of source galaxies
selected by our method. In this figure, we show small cutout images
around three massive lens galaxies at $z\sim0.25$. Red ellipses
indicate galaxies that have overlapping Kron ellipses. To predict the
\sn of $\Delta\Sigma$ measurements within $R_{\rm
  eq}$, we only use source galaxies with green ellipses.

\subsection{Impact of Ground-Based Photometric Redshifts on Source
  Counts}\label{subsec:photoz}

In addition to shape measurements, weak lensing also requires
photometric redshifts. The assignment of photometric redshifts to
galaxies in close-pair configurations is non-trivial. Photometric
redshifts are often derived from ground based imaging. For example,
both {\it Euclid} \citep[]{Laureijs2011} and {\it WFIRST} \citep[]{Spergel2013}
will need to complement their space-based imaging with ground based
photometry. Because ground based imaging typically has a PSF
of about one arcsecond, it may be difficult to derive
accurate photometry for source galaxies at $R_{\rm eq}$.

We use the COSMOS photo-z catalogue to quantify how many source galaxies
we lose due to this additional photometric redshift requirement. The
detection of sources in this catalogue was carried out on a combined
CFHT $i^*$ and Subaru $i^+$ image (with the original PSF of 0\farcs95)
\citep[]{Capak2007}. Galaxy colours were measured with a 3\farcs0
aperture after PSF homogenization.

We now investigate the conditions under which shape and photometric
redshift measurements fail in the COSMOS catalogues. Let us consider a
galaxy A with a close-by companion galaxy B. We would like to know
when shape and photo-z measurements typically fail for galaxy A as a
function of the distance from B and as a function of the brightness
ratio between galaxies A and B. To answer this question, we bin all of
the galaxies in our ACS catalogue by magnitude and refer to it as our
``primary'' sample. For every galaxy in this sample, we identify all
its neighbours in the entire ACS catalogue\footnote{These neighbours could
  themselves be a part of the primary sample.}. We use a criteria
based on Kron ellipses in order to scale distances between the primary
and its neighbours. Pairs of primary-neighbour galaxies whose Kron
ellipses overlap when their major and minor axes are scaled by a
factor $f_1<f<f_2$, are labelled $\sc{R}_{[f_1,f_2]}$. If there is
more than one overlapping neighbour, we use the brightest one (this
occurs in about $6$ to $10$ per cent of cases depending upon the value
of $f$).

\begin{figure*} \centering{
    \includegraphics[width=0.8\textwidth,keepaspectratio]{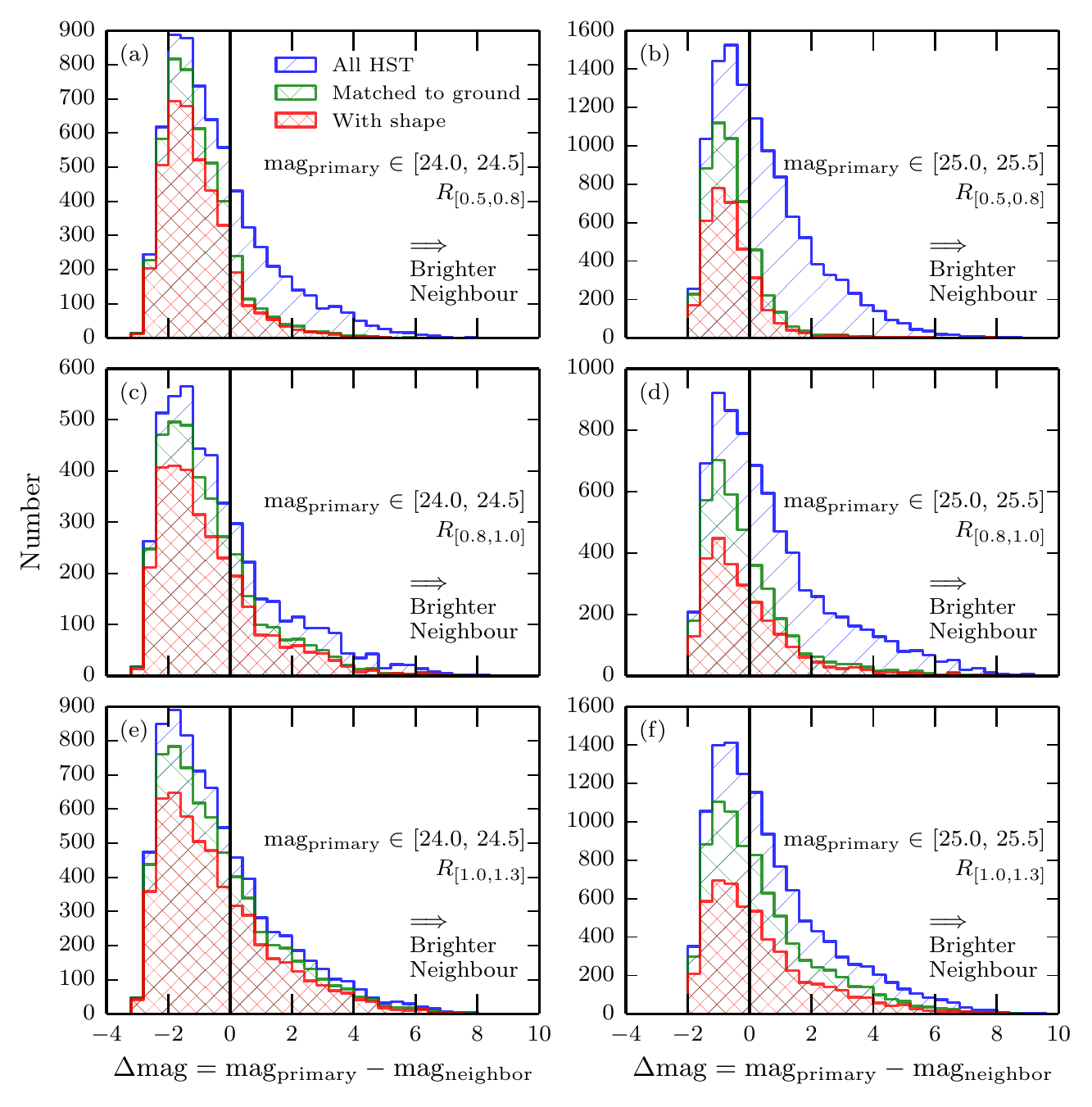}}
    \caption{ 
        Number of galaxies in COSMOS as a function of the magnitude
    difference (mag(primary galaxy) - mag(neighbour galaxy)) between
    galaxies in close-pair configurations.  Left panels: primary
    galaxies with magnitudes $\in [24.0, 24.5]$. The distance to the
    closest neighbour increases from top to bottom ($R_{[0.5, 0.8]}$
    is a small separation and $R_{[1.0, 1.3]}$ is a large
    separation). Blue histograms represents all galaxies in the COSMOS
    catalogue. Green histograms (``Matched to ground'') represent
    primary galaxies that have a photometric redshift measurement. Red
    histograms (``With shape'') represent primary galaxies that have a
    shape measurement and that pass our quality cuts as discussed in
    Section~\ref{subsec:weaklensingcatalog}.  Right panels: same as
    left panels but for primary galaxies with magnitudes $\in [25.0,
    25.5]$. This figure shows that both shape and photo-z measurements
    fail for source galaxies with bright companions. For example,
    panel d) shows that shape measurements fail
    for about 50 per cent of galaxies with magnitude $\in [25.0, 25.5]$ and
    which have a close-by companion of equal magnitude ($\Delta
    mag=0$). This failure rate increases sharply for brighter
    companions ($\Delta mag>0$). Panel e) however shows that at
    mag$\in [24.0, 24.5]$ and for $1.0<f<1.3$, most galaxies have both
    shape and photo-z measurements. A key point to note from this
    figure is that the red histograms are typically lower than the
    green: shape measurements in the COSMOS catalogue are a more
    stringent requirement compared to photo-zs.}
\label{fig:magdif_6panel_subaru}
\end{figure*}

Fig.~\ref{fig:magdif_6panel_subaru} shows how often shapes and
photo-zs are measured for primary galaxies as a function of the
distance and magnitude difference with the brightest neighbour. The
two columns correspond to different primary samples (the fainter primary
sample is on the right). The top row corresponds to galaxies in close
pairs where the separations are small ($R_{[0.5,0.8]}$). We
investigate close-pair galaxies with successively larger separations
in the middle ($R_{[0.8,1.0]}$) and the bottom row ($R_{[1.0,1.3]}$).
Fig.~\ref{fig:magdif_6panel_subaru} shows that shape and photo-z
measurements are more likely to fail for galaxies which have bright
companions. Comparing the left and the right hand panels shows that
this effect is more severe for faint galaxies. Comparing the different
rows it is clear that the effect becomes smaller as we consider
galaxies in close pairs but with larger separations.

In this test, we only consider {\it when} a galaxy has a photometric
redshift measurement, but we do not evaluate whether or not photo-zs
for galaxies with close by companions have a larger bias or a larger
fraction of catastrophic errors. In addition, here we also only
consider {\it when} a shape measurements has been possible, not
whether or these shape measurements are biased. Obviously, these are
aspects that need to be evaluated more critically in future work.
Bearing these caveats in mind, Fig.~\ref{fig:magdif_6panel_subaru}
shows that shape measurements in the COSMOS catalogue are a more
stringent requirement compared to photo-zs.

\subsection{Proximity Effects in the Number of Source Galaxies as a Function of Transverse Distance from Lenses}
\label{subsec:ns_radialdep}

\begin{figure*} \centering{
    \includegraphics[width=0.7\textwidth,keepaspectratio]{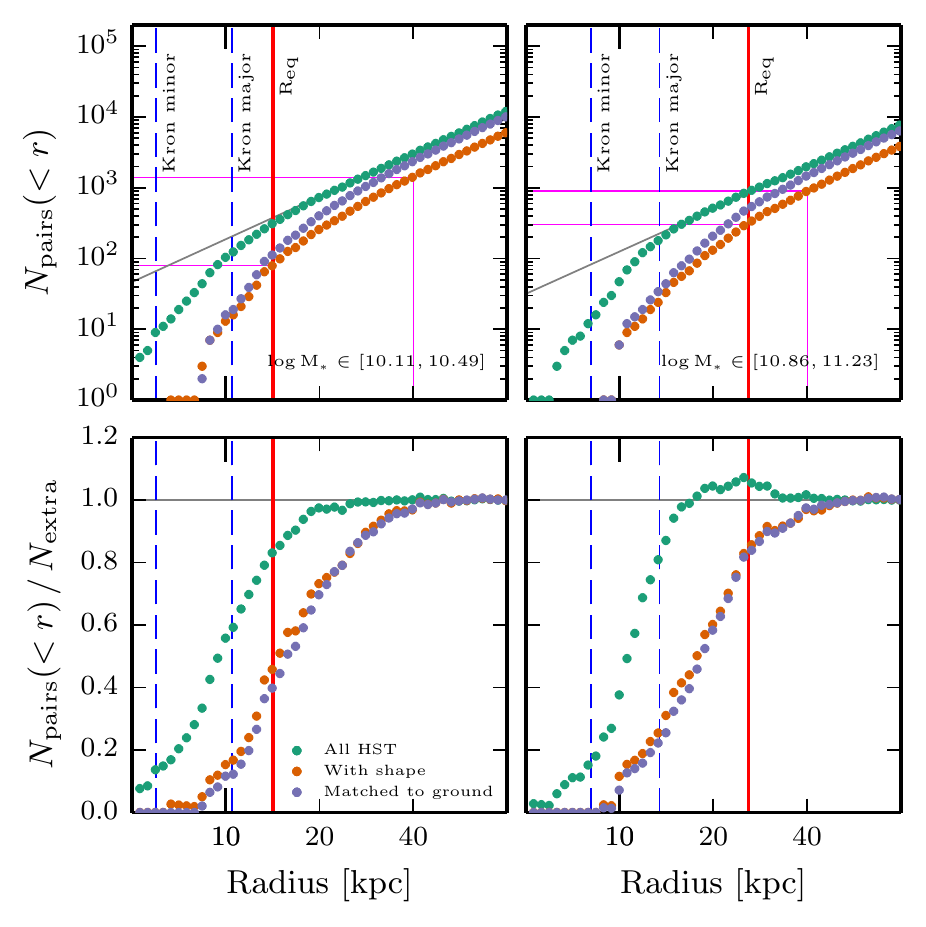}}
  \caption{Upper panels: cumulative number of lens-source pairs,
    $N_{\rm pairs}(<r)$, as a function of the transverse distance from
    lens galaxies at $z \in \left[0.62, 0.72\right]$. Left panels
    correspond to lens galaxies with $\log_{10} M_{*} \in \left[10.11,
      10.49\right]$. Right panels correspond to lens galaxies with
    $\left[10.86, 11.23\right]$. Blue dashed vertical lines indicate
    the average semi-major and semi-minor axis of Kron ellipses for
    lens galaxies. The thick red vertical line denotes $R_{\rm eq}$.
    Green dots indicate all galaxies from the COSMOS weak lensing
    catalogue with $i<26$.  Red dots represent galaxies with shape
    measurements. Blue dots represent galaxies that are matched to the
    ground-based photo-z catalogue. Note that no photo-z cuts
      have been applied to limit source galaxies to background
      galaxies (these cuts will be applied in Fig.~\ref{fig:num_radialdep_withphotoz}. 
      Magenta vertical thin line shows the location of 
    $r = 40$ kpc, and magenta horizontal thin lines respectively show
    $N_{\rm pairs}(<40 \mathrm{kcp})$ and $N_{\rm pairs}(<R_{\rm eq})$.
    Lower panels: $N_{\rm pairs}(<r)$
    divided by the expected number based on an extrapolation from
    larger scales ($N_{\rm extra}$). The lower panels show that
    proximity effects starts to influence the source galaxy counts at
    $r\sim40$ kpc (roughly $1$ to $5$ times larger than $R_{\rm eq}$)
    and photo-z matching
    and shape cuts reduce $N_{\rm pairs}(<r)$ by
    20-60 per cent at $r<R_{\rm eq}$. Note that we have not applied any cuts
    to removed overlapping galaxies, this will further reduce $N_{\rm pairs}(<r)$. 
}
\label{fig:num_radialdep_wophotoz}
\end{figure*}

We investigate how blending affects the number of lens-source pairs as
a function of transverse separation, $r$. The upper panels of
Fig.~\ref{fig:num_radialdep_wophotoz} show the cumulative number of
observed lens-source pairs, $N_{\rm pairs}(<r)$, in two different
stellar mass bins and at one fixed redshift bin. Symbols with
different colours correspond to all HST detected galaxies, those with
photometric redshifts, and those that pass our source galaxy
cuts. Note, however, that we have not yet imposed any proximity cuts
on the source catalogue -- galaxies with overlapping Kron ellipses are
still included at this stage. For all samples in this figure, we also
impose a magnitude cut at $i<26$. Note that we also do not yet impose
a photo-z cut 
to separate foreground and background objects -- we simply
consider all pairs along the line-of-sight.

Fig.~\ref{fig:num_radialdep_wophotoz} shows that the overall number
of lens-source pairs at large separations ($\sim 40$ kpc) decreases as
we consider higher mass lens galaxies (indicated by the magenta vertical
thin line). However, high mass lens galaxies have a larger number of
lens-source pairs within \req (indicated by the magenta horizontal thin
line). This is consistent with our expectations from
Section~\ref{subsec:evolution_req}: both \req and \Aeff are larger for
more massive galaxies. Because massive galaxies have a larger
effective area where source galaxies can be found, they are more
suited for detecting the small scale weak lensing signal.

The bottom panels of Fig.~\ref{fig:num_radialdep_wophotoz} show the
ratio of $N_{\rm pairs}(<r)$ to the number of pairs expected from a
simple power law extrapolation (with slope $-2$) of $N_{\rm pairs}(<r)$
from larger scales (thin grey line). These panels show that although a
number of galaxies are detected at $r<R_{\rm eq}$, our photo-z and
shape measurement requirements bring these numbers down
significantly. Also, as discussed in the previous section,
Fig.~\ref{fig:num_radialdep_wophotoz} shows that we lose more
source galaxies due to the shape measurement requirement than
the photo-z requirement.
Note that all HST pairs in the higher mass lens bin
outnumber the large-scale extrapolation on scales $\sim20\mathrm{kpc}$.
This bump feature indicates that 
galaxies are clustered with lens samples along the line-of-sight.
These galaxies can be misinterpreted as source galaxies in our \sn predictions
due to the photo-z errors.
In Section~\ref{subsec:predict_onebin},
we will discuss such potential confusion in source selection due to
correlated galaxies.

\begin{figure*} \centering{
    \includegraphics[width=0.7\textwidth,keepaspectratio]{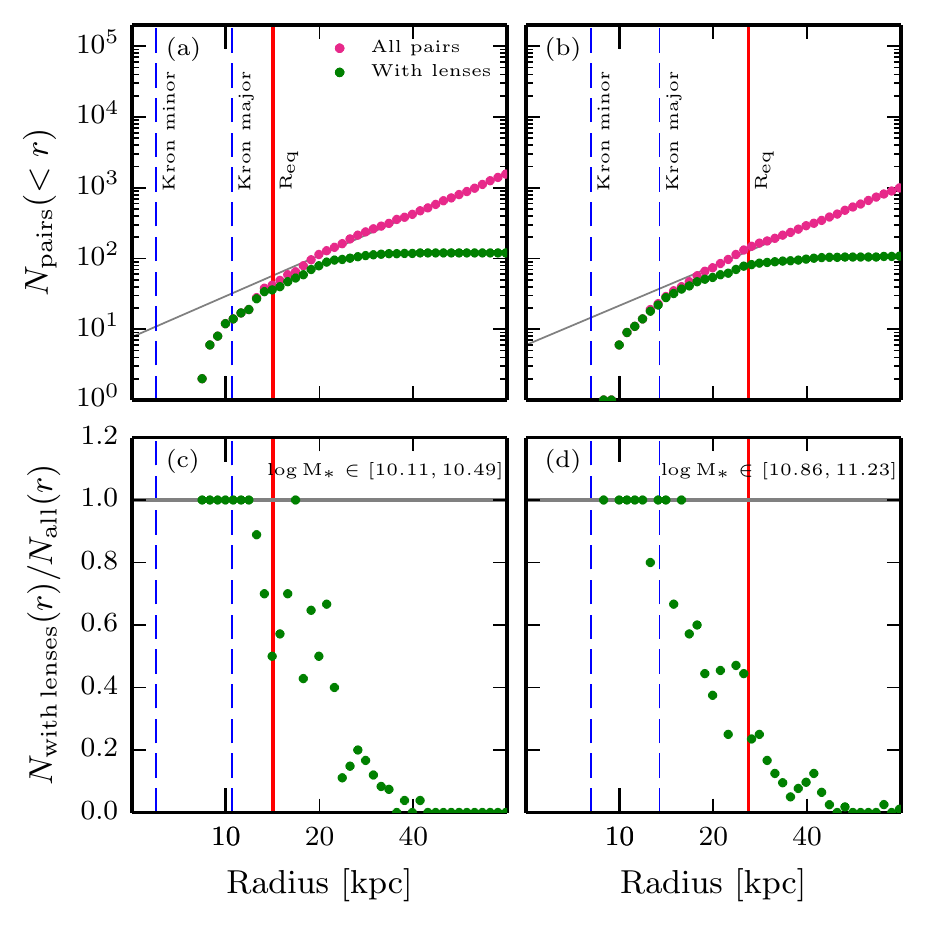}}
  \caption{Similar to Fig.~\ref{fig:num_radialdep_wophotoz} but here
    we only show galaxies which have a Kron ellipse that overlaps with
    a neighbouring galaxy ($f=1$). Upper panels: cumulative number of
    galaxies identified as close pairs as a function of the transverse
    distance from lens galaxies. We distinguish overlaps that occur
    between a source and a lens from the sample under consideration
    from overlaps that occur between a source and some other non-lens
    galaxy. Magenta dots indicate the cumulative number of all source
    galaxies that are flagged are overlaps. Green dots represent
    source galaxies that have a Kron ellipse that overlaps with the
    Kron ellipse of a lens galaxy.  
    The overall amplitude of $N_{\rm pairs}(<r)$ on larger radial scales
    is dominated by overlap between source galaxies that are spatially
    uncorrelated with the lens sample. On the other hand,
    the radial dependence of $N_{\rm pairs}(<r)$ on smaller radial scales
    is mainly driven by overlap with the primary lens sample and with
    galaxies that are correlated with the lens galaxies. 
    Note that no photo-z cuts
    have been applied to limit source galaxies to background
    galaxies (these cuts will be applied in Fig.~\ref{fig:num_radialdep_withphotoz}).
    Lower panels: fraction of source
    galaxies flagged as overlaps and for which the overlap occurs with
    a lens galaxy from the sample under consideration. Panel c)
    shows that for the lower stellar mass bins,
    overlaps mainly occur at \req because source galaxies tend to
    overlap with galaxies from the primary lens sample. Panel d) on
    the other hand, shows that for the higher stellar mass bins,
    only $\sim$30 per cent of the source galaxies flagged as overlaps are
    colliding with the primary lens sample. The remaining $\sim$70 per cent
    overlap with either galaxies that are spatially correlated with the lens sample
    or with galaxies that are uncorrelated with the lens galaxies.}
\label{fig:num_radialdep_onlyoverlap_wophotoz}
\end{figure*}

In addition to these effects, galaxies will also be rejected from our
source catalogue due to the proximity cuts defined in
Section~\ref{subsec:proximitycuts}.  We are interested to know if
source galaxies are flagged as overlaps at \req because they overlap
with the lens galaxy under consideration, or because they overlap 
with non-lens galaxies.. Fig.~\ref{fig:num_radialdep_onlyoverlap_wophotoz} quantifies
the relative importance of the two effects. For low stellar mass
galaxies, we find that the majority of source galaxies lost within
\req are due to overlap with the lens galaxy itself. At higher stellar
masses we find that only $\sim$30 per cent of the source galaxies flagged as
overlaps are colliding with the primary lens sample. The remaining
$\sim$70 per cent overlap with 
either nearby galaxies 
that correlate with the primary lens sample,
or with galaxies that are spatially uncorrelated with the lens sample.
Again in this figure, we consider all pairs along the line-of-sight
without the photo-z cut.

\begin{figure*} \centering{
    \includegraphics[width=0.7\textwidth,keepaspectratio]{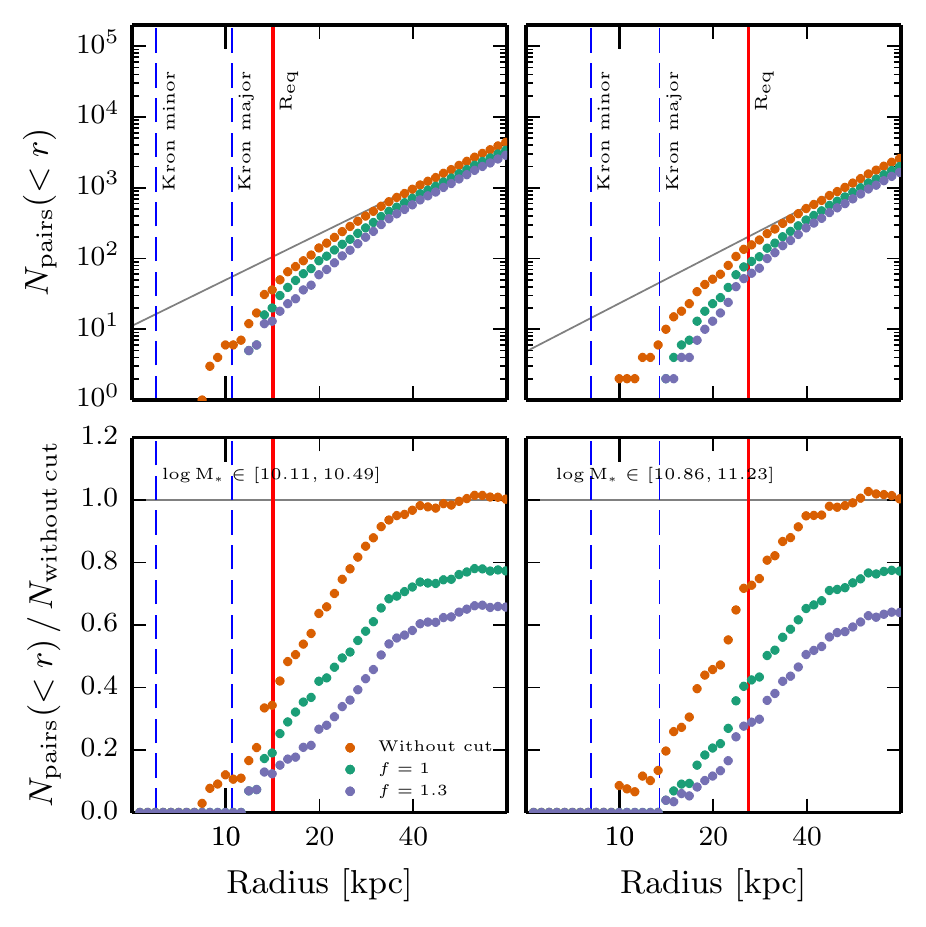}}
  \caption{Upper panels: similar to
    Fig.~\ref{fig:num_radialdep_wophotoz} but here we show only source
    galaxies that pass also the photo-z cut. Red dots represent all source galaxies with both shape
    and photo-z measurements. Green dots represent only source galaxies
    that pass our proximity cut with $f=1$.  Similarly, blue dots
    represent only source galaxies that pass our proximity cut with
    $f=1.3$.  Lower panels: $N_{\rm pairs}(<r)$ divided by $N_{\rm
      without \, cut}$ where $N_{\rm without \, cut}$ is the
    extrapolation from larger scales of the source counts without any
    proximity cuts (red points). With $f=1$, the overall source
    density is reduced by 20 per cent. With $f=1.3$, the overall source
    density is reduced by 30 per cent (also see Table
    \ref{table_ngal}). Proximity effects start to impact $N_{\rm
      pairs}(<r)$ at $40$ kpc. At the equality radius, the source
  density is reduced by $60$--$80$ per cent for $f=1$.}
\label{fig:num_radialdep_withphotoz}
\end{figure*}

Finally, to select only source galaxies located at higher
redshifts than lens galaxies, we impose a photo-z cut
$z_{\rm s}>z_{\rm l, max}$ where $z_{\rm l, max}$
represents the upper redshift limit of the lens bin under
consideration $[z_{\rm l, min}, z_{\rm l,max}]$. 
The combination of the photo-z cut, the shape
measurement cut and the proximity cut results in the cumulative number
distribution of lens-source pairs shown in
Fig.~\ref{fig:num_radialdep_withphotoz}.  Symbols with different
colours correspond to source galaxies that pass only photo-z cut and
shape measurement cut (in red), those that pass also proximity cut
with $f=1$ (in green), and those that pass also proximity cut with
$=1.3$ (in blue).  In the fiducial case ($f=1$), the fraction of
source galaxies within $R_{\rm eq}$ that pass the photo-z cut, but are
rejected due to a combination of the shape measurement cut and the
proximity cut is over $80$ per cent.  We will use the $N_{\rm
  pairs}(<R_{\rm eq})$ in order to estimate the error on the weak
lensing measurement within $R_{\rm eq}$ for every stellar mass and
redshift bin in our sample. We compute the weak lensing signal at the
average separation between the lens-source pairs using
Equations~\ref{eq:delsig_dmgx} and \ref{eq:delsig_stellar}.

\subsection{$N_{\rm pairs}(<R_{\rm eq})$ : Number of Lens-Source Pairs
  Within the Equality Radius}
\label{subsec:nlens_source_pairs}
We compute the total number of source galaxies that pass
the photo-z cut, the shape measurement cut, the proximity cut, 
and which have $r<R_{\rm eq}$ for each of our lens bins. 
The values for $N_{\rm pairs}(<R_{\rm eq})$ for $f=1$
are given in Table \ref{table_npair}.
As can be seen from Table~\ref{table_npair}, 
the overall number of lens-source pairs for any
given bin is small, only of order $\sim100$.
The difference between Tables~\ref{table_nl} and \ref{table_npair}
agrees with the results our conclusions from Section~\ref{subsec:ns_radialdep} -- low-mass lens galaxies have almost no source galaxies within $R_{\rm eq}$
due to the small size of $R_{\rm eq}$.

\begin{table*}
    \begin{tabular}{c|c|c|c|c|c|c|}
       \hline
       \hline
        Redshift & \multicolumn{6}{|c|}{$\log_{10} M_{*} [{\rm M_{\odot}}]$}\\
& $[$ 9.37 , 9.74 $]$& $[$ 9.74 , 10.11 $]$& $[$ 10.11 , 10.49 $]$& $[$ 10.49 , 10.86 $]$& $[$ 10.86 , 11.23 $]$& $[$ 11.23 , 11.60 $]$\\
\hline 
 \hline
$[$ 0.30 , 0.50 $]$ & 1 & 14 & 66 & 197 & 178 & 51 \\ 
$[$ 0.50 , 0.62 $]$ & 0 & 1 & 21 & 70 & 84 & 29 \\ 
$[$ 0.62 , 0.72 $]$  & $\times$ & 1 & 21 & 60 & 87 & 48 \\ 
$[$ 0.72 , 0.80 $]$  & $\times$ & 1 & 12 & 37 & 52 & 24 \\ 
$[$ 0.80 , 0.87 $]$  & $\times$  & $\times$ & 17 & 55 & 91 & 39 \\ 
$[$ 0.87 , 0.94 $]$  & $\times$  & $\times$ & 6 & 31 & 43 & 27 \\ 
$[$ 0.94 , 1.00 $]$  & $\times$  & $\times$ & 7 & 40 & 57 & 25 \\ 
 \hline 

    \end{tabular}
    \caption{The number of lens-source pairs for $r<R_{\rm eq}$ for
        each of our lens bins. Source galaxies which have Kron ellipses
        that overlap with any nearby galaxy ($f$=1) are rejected from
        the source catalogue before computing these numbers. The
        photo-z cut to seperate background and foreground source galaxies
        has also been applied before compute these numbers.}
    \label{table_npair}
\end{table*}

\subsection{Error on $\Delta\Sigma$ and Signal to Noise Ratio}
\label{subsec:err_delsig}
Following \citet[]{Leauthaud2007}, the error on the shear for each
source galaxy is estimated as a combination of intrinsic shape noise ($\sigma_{{\rm int}} = 0.27$)
and shape measurement error:

\begin{equation}
    \sigma_{\tilde{\gamma}}^{2}= \sigma_{{\rm int}}^{2} + \sigma_{{\rm meas}}^{2}\,.
\end{equation}
The optimal weight for each source galaxy is given by
\begin{equation}
    w_{i} = \frac{1}{\left( \Sigma_{{\rm crit}, i}\, \sigma_{\tilde{\gamma}, i}
\right)^2} \,,
\end{equation}
and the estimated error on $\Delta \Sigma$ in the COSMOS survey is given by
\begin{equation}
        \sigma^{\rm COSMOS}_{_{\Delta \Sigma}} = \left[ \sum\limits_{i}^{N_{\rm
        pairs}(<R_{\rm eq})} w_{i} \right]^{-1/2} \,, \label{eq:weight_def}
\end{equation}
The sum runs over all lens-source pairs with $r<R_{\rm eq}$ in a given
stellar mass and redshift bin.

To scale our error estimate to survey areas larger than COSMOS, we
assume that the distribution of source galaxy redshift and shape
measurement errors in COSMOS are representative and we simply scale
the estimated errors according to
\begin{equation}
        \sigma^{\rm survey}_{_{\Delta \Sigma}}=\sigma^{\rm COSMOS}_{_{\Delta
        \Sigma}}\sqrt{ \frac{A_{\rm COSMOS}}{A_{\rm survey}} }\,.
        \label{eq:survey_scaling}
\end{equation}
where $A_{\rm COSMOS}=1.64$ deg$^2$.
Finally, combining $\Delta \Sigma$ and $\sigma_{\Delta \Sigma}$
(Equations~\ref{eq:delsig_dmbar} and \ref{eq:weight_def}), the
predicted \sn in each stellar mass and redshift bin
for COSMOS is simply

\begin{equation}
    {\rm S/N} = \frac{\Delta \Sigma}{\sigma_{\Delta \Sigma}^{}} \,. \label{eq:sn}
\end{equation}

\subsection{Predictions for COSMOS}
\label{subsec:predictions_for_cosmos}
Using Equations~\ref{eq:weight_def} and \ref{eq:sn}, we compute the expected \sn 
for one weak lensing data point at $r<R_{\rm eq}$ in the COSMOS ACS catalogue.
The results are shown in Fig.~\ref{fig:sn_predicted_cosmos}.
We find that the \sn for this type of measurement is maximized
for massive galaxies at low redshifts, in agreement with our intuitive
expectation from Section~\ref{subsec:weakshear_regime}.
We also show the effects of
making more or less conservative source galaxy selections by varying
our $f$ factor. A larger value of $f$ means that we reject more source
galaxies in close-pair configurations. We find that varying $f$
between 0.8 and 1.3 only has a relatively minor impact on the
predicted \sn. This implies that a more conservative selection of
sources galaxies does not necessarily have a strong impact on the
\sn.


\section{Predictions for {\it Euclid} and {\it WFIRST}}
\label{sec:predict_for_space}
In this section, we present our predictions for the \sn 
for small scale weak lensing measurements from future space-based
surveys {\it Euclid} and {\it WFIRST}.

\subsection{Next Generation Weak Lensing Space Based Surveys}

{\it Euclid} is a space mission under development for an expected launch in 2020
\citep[]{Laureijs2011}. {\it Euclid} consists of a $1.2$ m Korsch telescope
and two instruments, the VIS (visible imager) and NISP (Near Infrared
Spectrometer and Photometer) and is designed to study the properties
of dark energy and dark matter and to search for evidence of modified
gravity by using weak gravitational lensing and galaxy clustering.
{\it Euclid} will perform a 6 year imaging and spectroscopic survey over the
lowest background $15000$ deg$^2$ of the extragalactic sky.
Visible imaging in a single wide \textit{riz} filter will provide
shapes for about $1.6$ billion galaxies; near infrared imaging in
$Y,J,H$ bands combined with ground-based photometry will enable high
precision photometric redshifts for source galaxies. {\it Euclid} will
have a $\sim$0.2 arcsecond PSF and will measure galaxy shapes with a
source density of $\sim$30 galaxies per arcmin$^{2}$.

The {\it Wide Field Infrared Survey Telescope} ({\it WFIRST}) is a NASA mission
under study for possible launch in $\sim2024$ \citep[]{Spergel2013}.
{\it WFIRST} consists of $2.4$m telescope a Wide Field Imager with 18
4k$\times$4k near infrared (NIR) detectors with $0.1$ arcsecond
pixels, an integral field unit (IFU) spectrograph, and an exoplanet
coronagraph.  {\it WFIRST} is designed to perform NIR surveys for a wide
range of astrophysics goals, including weak lensing. {\it WFIRST}'s weak
lensing survey will cover $2400$ deg$^2$ and make galaxy shape
measurements in $3$ NIR bands of about $500$ million galaxies over the
course of two years in {\it WFIRST}'s primary mission of 6 years. {\it WFIRST}
will have a PSF of 0.1 arcseconds (before pixellization) at 1 micron
and about 0.2 arcseconds at 2 microns and will measure galaxies shapes
with a source density of $\sim$54 galaxies per arcmin$^{2}$.

\subsection{Predictions for one bin at $r<R_{\rm eq}$}
\label{subsec:predict_onebin}
To make our predictions for {\it Euclid}, we assume a source density of 30
galaxies per arcmin$^{2}$ over $15000$ deg$^2$. To mimic this
source density, we simply apply a detection signal-to-noise cut on our
COSMOS catalogue. After applying a signal-to-noise cut  that yields 30
galaxies per arcmin$^{2}$, we rederive $N_{\rm pairs}^{EUCLID}(<r)$ for {\it Euclid}.

{\it WFIRST} is expected to achieve a source density of $\sim$54 galaxies
per arcmin$^{2}$ over $2400$ deg$^2$. The COSMOS source density
with reliable shape measurements and photo-zs 
(Section~\ref{subsec:weaklensingcatalog})
is only of order $\sim$45 galaxies per arcmin$^{2}$,
thus we cannot directly generate predictions for the expected {\it WFIRST}
source density. Instead, for {\it WFIRST}, we will use 45 galaxies per
arcmin$^{2}$. Our estimates for {\it WFIRST} will be conservative in this
regard.

As mentioned earlier in Section~\ref{subsec:err_delsig}, our
predictions make several simplifying assumptions. First, we use the
same shape noise as COSMOS (we do not attempt to adjust the
measurement error component of the shape noise). Secondly, we use
photometric redshifts from COSMOS, ignoring the fact for example, that
{\it WFIRST} will detect galaxies in redder bands, and thus a somewhat
higher redshifts. Finally, we also ignore the extra effects of
smearing by a larger PSF ($\sim$ a factor of $1$--$2$ larger compared to
COSMOS) on $N_{\rm pairs}(r)$, the number of source galaxies as a
function of transverse separation. A more detailed study should
account for such effects, but our goal here is simply to generate a
first order prediction for the expected \sn.

\begin{figure*} \centering{
\includegraphics[scale=1.1]{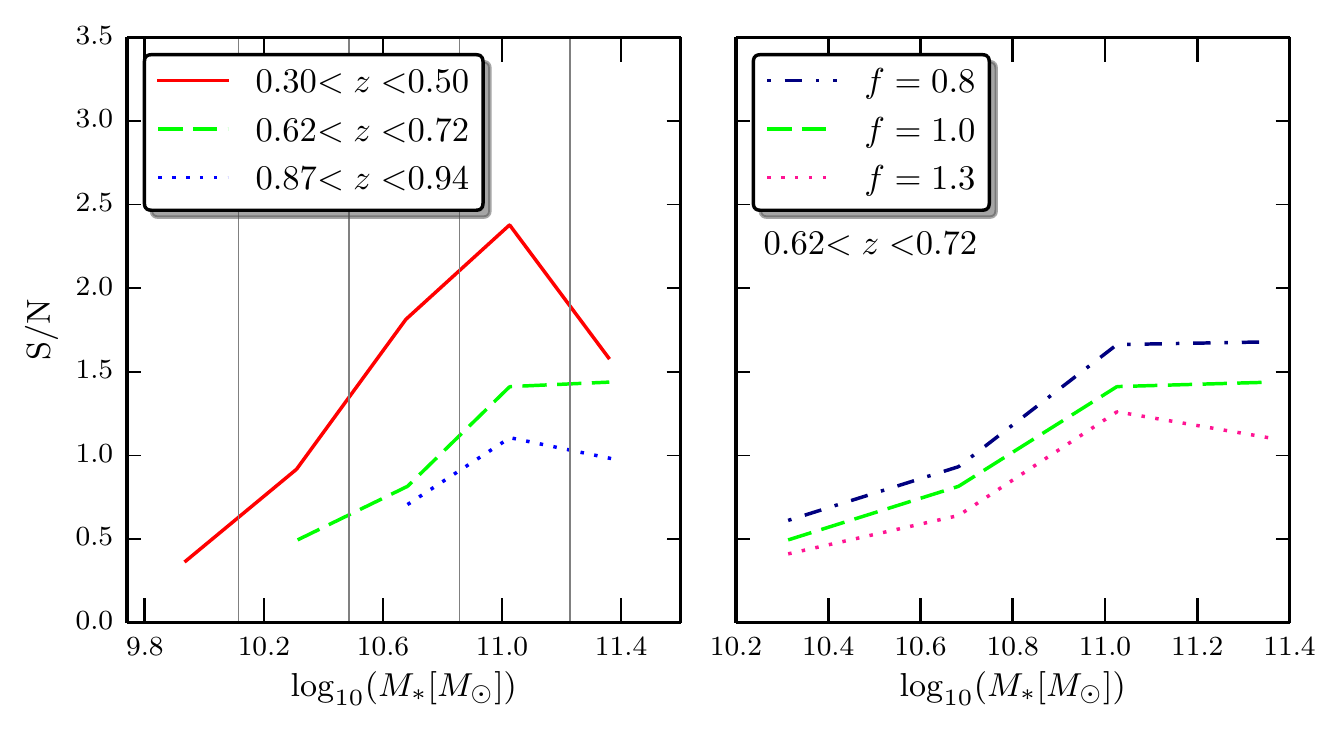}}
\caption{
  Predicted \sn for $\Delta\Sigma$ measured
  within $R_{\rm eq}$ as a function of lens stellar mass and
  redshift in the COSMOS field. Left: predicted \sn
  for three lens redshift bins
  assuming our fiducial close-pair cut with $f=1$. Grey vertical lines
  represent our lens stellar mass bins.  Note that the \sn
  with $z \in [0.3, 0.5]$ decreases rapidly at the highest stellar
  mass bin because the number of lens-source pairs decrease at high
  mass end (See Table~\ref{table_npair}). Right:
  expected \sn when
  we make more or less conservative cuts on source galaxies in close
  pair configurations (only the middle redshift bin is shown here). A
  larger value of $f$ corresponds to a more conservative selection of
  source galaxies. Varying $f$ between 0.8 and 1.3 only has a
  relatively minor impact on the predicted \sn.}
\label{fig:sn_predicted_cosmos}
\end{figure*}

\begin{figure*} \centering{
    \includegraphics[scale=1.1]{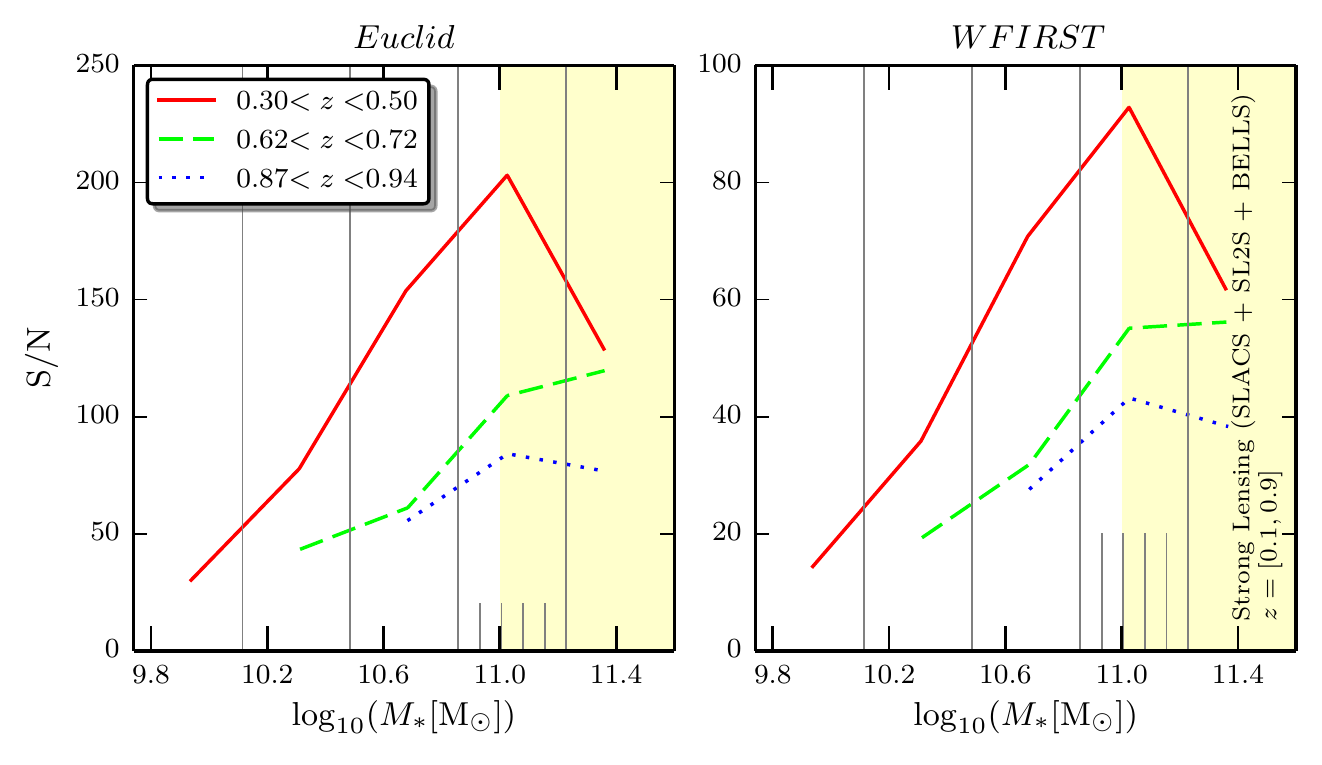}}
  \caption{Similar to Fig.~\ref{fig:sn_predicted_cosmos}, but
      for future surveys -- the left panel is for {\it Euclid} and the
      right panel is for {\it WFIRST}.  Again, we assume our fiducial
      close-pair cut with $f=1$ and grey vertical lines indicate our
      lens stellar mass bins.  Short grey thin lines represent the
      division of one the stellar mass bins into 5 smaller bins of
      width $0.074$ dex. Our prediction signal in Figs.~\ref{fig:delsig_constrained_wfirst} 
      and \ref{fig:delsig_constrained_euclid}
    will adopt this finer binning scheme. Note that the
    \sn with $z \in [0.3, 0.5]$ decreases rapidly in the highest
    stellar mass bin due to smaller number of lens-source pairs (also
    seen in Fig.~\ref{fig:sn_predicted_cosmos}).  The yellow
    shaded region with $\log_{10}(M_*)[\msun] \geq 11.0$ indicates the
    typical stellar mass range for strong lensing samples from the SLACS, SL2S, and BELLS surveys with $z \in
    [0.1,0.9]$.  Although {\it WFIRST} will have a higher source
    density, its smaller area results in smaller \sn compared with {\it Euclid}.}
  \label{fig:sn_predicted_future}
\end{figure*}

We compute $N_{\rm pairs}^{EUCLID}(<r)$ for {\it Euclid} and use $N_{\rm
  pairs}^{COSMOS}(<r)$ for {\it WFIRST}. The expected \sn for one
  weak lensing data point at $r<R_{\rm eq}$ is then computed using
Equations~\ref{eq:survey_scaling} and \ref{eq:sn}. The results are
shown in Fig.~\ref{fig:sn_predicted_future}. 
Similarly to Fig.~\ref{fig:sn_predicted_cosmos},
we observe that the \sn for both {\it Euclid} and {\it WFIRST} reaches
its maximum for massive galaxies at lower redshifts.
Fig.~\ref{fig:sn_predicted_future} 
indicates that
{\it Euclid} and {\it WFIRST} may detect $\Delta\Sigma$ on very small radial
scales with very high \sn greater than 20 for lens
galaxies with $\log_{10}(M_*)>10.4$.
Note that this is lower mass than the mass probed by strong lensing,
which is typically $\log_{10}(M_*)>11.0$ 
\citep[\eg,][]{Auger2010a,Oguri2014,Sonnenfeld2014}.

\begin{figure*}\centering{
    \includegraphics[scale=1.1]{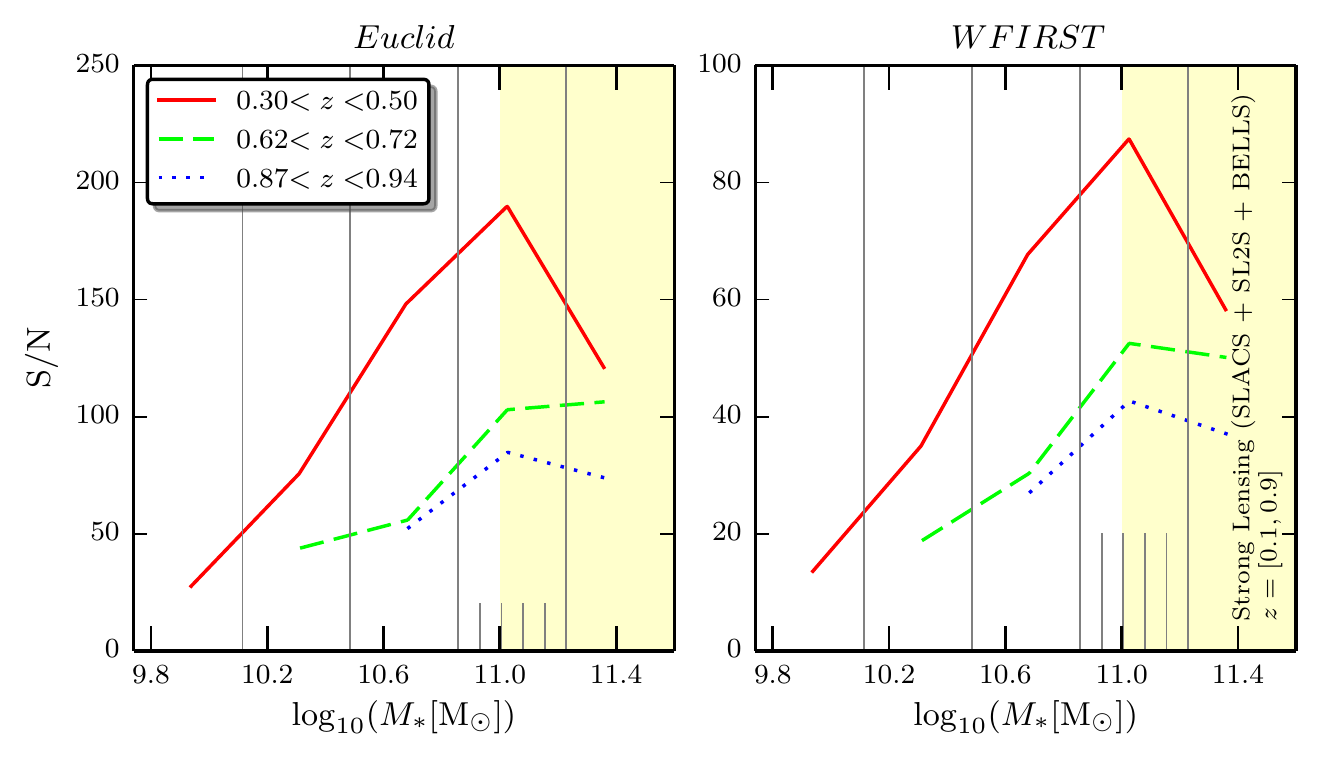}}
    \caption{Similar to Fig.~\ref{fig:sn_predicted_future}
        but with a stringent photo-z cut applied (see Equation~\ref{eq:buffer_zred}).
        Compared with Fig.~\ref{fig:sn_predicted_future},
        the \sn at the higher mass end decreases up to $10$ per cent but 
        the overall dependence on the stellar mass and the redshift of lens galaxies does not change.
        Therefore, the confusion in source selection due to galaxies that are
        correlated with lens samples can be mitigated by such photo-z cuts.
        This may reduce \sn by a few per cent, but does not alter our main
    conclusion -– massive galaxies at lower redshifts yield higher S/N.
    }
    \label{fig:sn_predicted_future_bufferredshift}
\end{figure*}

Until now, all analyses presented in this section involve the photo-z cut, which
uses best-fit photo-z on both lens and source samples.
However, as we observed in Fig.~\ref{fig:num_radialdep_wophotoz},
the confusion in source selection due to photo-z errors
might be buried in our analyses even after the close-pair cut is applied 
(\eg, galaxies that are correlated with lens samples).
To investigate such possibility, we employed several stringent photo-z cuts
to separate lens and source galaxies and re-compute the predicted \sn.
For example, to remove galaxies that are correlated with lens samples
but have $z>z_{\rm l,max}$,
we applied an additional buffer redshift to the maximum photo-z in each lens bin as
\begin{equation}
    z_{\rm s} > z_{\rm l,max} + 0.1 (1+z_{\rm l,max}) \,.
    \label{eq:buffer_zred}
\end{equation}
The buffer of $0.1$ is motivated by results from \citet[]{Ilbert2009},
who report a one-sigma error on the best-fit photo-z
of about $\sigma_{z}/(1+z) = 0.07$ for objects with $i^+\sim25.5$,
which is more conservative than the expected errors 
in {\it Euclid} and {\it WFIRST} (see \citealt{Laureijs2011} 
and \citealt{Spergel2013} respectively).
The resultant \sn is shown in Fig.~\ref{fig:sn_predicted_future_bufferredshift}.
The predicted \sn decrease up to $10$ per cent
at the higher mass end but the stringent cut (Equation~\ref{eq:buffer_zred}) 
does not alter the overall trend -- the \sn is maximized at
massive galaxies at lower redshifts.

\subsection{Predictions for radial bins}
\begin{figure*} \centering{
    \includegraphics[scale=2.0]{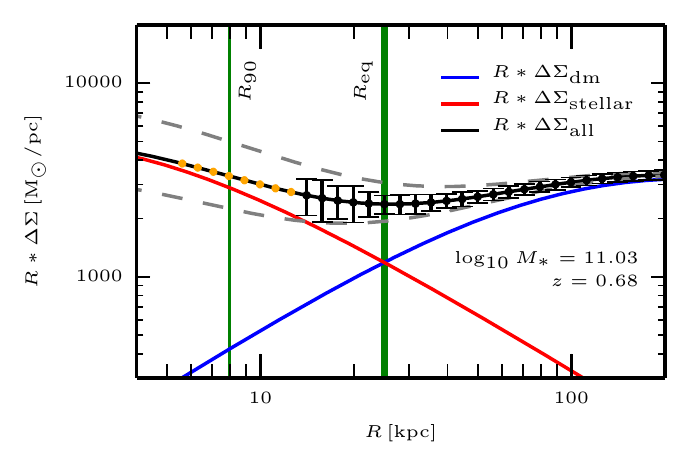}}
  \caption{Predicted \sn on the $\Delta\Sigma(r)$ profile of lens
      galaxies with $\log_{10}(M*)=11$ for {\it WFIRST}. Note that the vertical
    scale is multiplied by $R$ to highlight the error bars. The dark
    matter profile ($\Delta\Sigma_{\rm dm}$) is shown by the blue line
    and the stellar component ($\Delta\Sigma_{{\rm stellar}}$) is
    shown by the red line. The black line represents the sum of
    $\Delta\Sigma_{\rm dm}$ and $\Delta\Sigma_{\rm stellar}$.  The
    green thin vertical line indicates the radius that encompasses
    $90$ per cent of the flux of the lens galaxies. The green thick vertical
    line shows $R_{\rm eq}$.  Grey dashed lines show $\Delta\Sigma$
    when $M_*$ is varied by $\pm 0.2$ dex (roughly corresponding to
    the current uncertainty on the IMF). Yellow points indicate radial
    bins where there are no source galaxies in our COSMOS catalogue.}
\label{fig:delsig_constrained_wfirst}
\end{figure*}

\begin{figure*} \centering{
\includegraphics[scale=2.0]{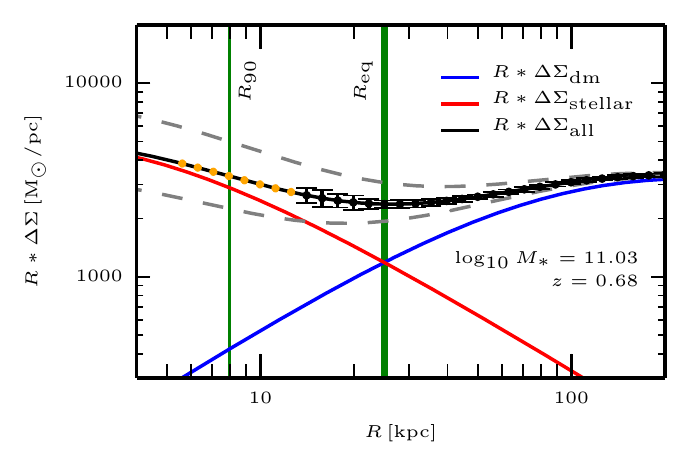}}
\caption{Same as Fig.~\ref{fig:delsig_constrained_wfirst} but for
  {\it Euclid}. {\it Euclid} will have a lower source density compared to {\it WFIRST}
  but will observe an area $6$ times larger ($15000$ deg$^2$ versus
  $2500$ deg$^2$). This results in a factor of 2 in the expected \sn
  for {\it Euclid} compared to {\it WFIRST}. Note, however, that we have neglected
  other effects (e.g., the PSF size of EUCLID is twice as large as
  {\it WFIRST}) which will probably affect these predictions at a similar
  level.}
\label{fig:delsig_constrained_euclid}
\end{figure*}

Figs.~\ref{fig:sn_predicted_cosmos} and \ref{fig:sn_predicted_future}
show the expected overall \sn
for one bin at $r<R_{\rm eq}$. We now consider the expected \sn
in finer radial
bins. Figs.~\ref{fig:delsig_constrained_wfirst} and
\ref{fig:delsig_constrained_euclid} show the predicted errors on
$\Delta \Sigma$ for lens galaxies with the mean stellar mass of
$10^{11.03}$\msun with the bin width of $0.074$ dex ($5$ times
smaller than our bins in Table~\ref{table_npair}) at a mean redshift of
$z=0.68$. The radial binning scheme is arbitrary -- we opt to make
$451$ bins with logarithmically equal width from $0.013$ to $419$
kpc. An NFW profile is assumed for the dark matter component and a
Hernquist profile is assumed for the stellar component. This
prediction uses source galaxies that pass all the quality cuts
introduced in Section~\ref{subsec:weaklensingcatalog} including our
fiducial close pair cut with $f=1.0$.

As a reference, in Figs.~\ref{fig:delsig_constrained_wfirst} and
\ref{fig:delsig_constrained_euclid} we show how $\Delta \Sigma$ varies
when the stellar mass of the lens sample varies by $\pm 0.2$ dex (a
factor of $1.5$). As can be seen from these figures, both {\it WFIRST} and
{\it Euclid} should be able to tightly constrain a combination of the inner
dark matter slope and the total stellar mass of galaxies down to
$\log(M_*)>10.4$.

\section{Summary and Conclusions}
\label{sec:summary_and_conclusions}

Current constraints on dark matter density profiles from weak lensing
are typically limited to radial scales greater than $r\sim50$--$100$
kpc. Below this scale, there is a paucity of sources (background)
galaxies due to poor image quality (``seeing''), complicating effects
such as isophotal blending that inhibits shape measurements, and a
lack of source galaxies simply due to the relatively small areas
covered by current lensing surveys. Pushing weak lensing measurements
down to smaller radial scales, however, would open up the exciting
possibility of probing the very inner regions of galaxy/halo density
profiles.  Such measurements are currently only possible for strongly
lensed galaxies which are rare, occur mainly for massive galaxies, and
may also represent a biased sample. Weak lensing on these small scales
would extend constraints to a much wider range of redshifts, stellar
masses, and to an unbiased selection of galaxy types.

In this paper, we use a weak lensing catalogue from the COSMOS survey to
investigate the expected \sn of stacked weak lensing
measurements on radial scales of only a few tens of kpc. On these
scales, in addition to dark matter, the weak lensing signal is sensitive to the baryonic
mass of the host galaxy. Thus, future weak lensing measurements will
offer the exciting possibility of {\em directly} measuring
mass-to-light ratios and providing independent constraints on the Stellar 
Initial Mass Function (IMF) and on the interplay between baryons and
dark matter.

We introduce the ``Equality radius'', $R_{\rm eq}$, as the radius
where the dark matter component and the stellar component contribute
equally to the weak lensing observable, $\Delta\Sigma$. Using a
stellar-to-halo mass relation that has been calibrated for COSMOS
data, we compute the evolution of $R_{\rm eq}$ as a function of lens
stellar mass and redshift. We show that $R_{\rm eq}$ is of order $7$
kpc for $M_*=10^{9.6}$\msun and $34$ kpc for $M_*=10^{11.4}$\msun at
$z=0.4$.  For lens galaxies with $M_*>10^{10.3}$\msun, this equality
radius is about a factor of two larger than the size of the Kron
ellipse which encompasses roughly $90$ per cent of the light. This leaves a
very narrow window (width of order $10$ kpc) with which to measure
weak lensing signals on scales where the stellar component dominates
the lensing signal. The area of this window (per lens galaxy) $A_{\rm
eff}$ varies between $2\times10^{-4}$ arcminute$^2$ and $0.3$ arcminute$^2$.
We show that $A_{\rm eff}$ is maximal for high
mass galaxies at low redshifts due to the fact that a) $R_{\rm eq}$ is
large for massive galaxies and b) a fixed value of $R_{\rm eq}$ in
physical units corresponds to a larger angular size at low redshifts.

Using COSMOS, we calculate the number of lens-source pairs as a
function of transverse separation $N_{\rm pairs}(r)$ down to $r=5$
kpc. We show that in the COSMOS catalogue, more galaxies in close-pair
configurations are rejected because of cuts related to the quality of
{\it shape} measurements rather than due to cuts related to the
availability of a {\it photo-z} measurement. This test is simplistic
in the sense that we only consider the availability of a shape or a
photo-z measurement -- not the quality of these measurements. This
question obviously needs to be studied in greater detail, but this
simple tests suggests that shape measurements rather than photo-z
measurements may be the limiting factor for these types of studies.

We investigate how blending and proximity affect the source counts 
as a function of transverse distance from the lens sample. 
We show that within $R_{\rm eq}$, the number of source galaxies is reduced
by $\sim20$ per cent at $M_{*}=10^{11.0}$\msun and by $60$ per cent at $M_{*}=10^{10.3}$\msun due to blending.
This sharp decrease in the number of source
galaxies on small radial scales is due to the effects of
masking/blending by foreground galaxies \citep[]{Simet2014}. We
quantify how often blends occurring at $R_{\rm eq}$ are due to the lens
sample or to other (non lens) galaxies that are source galaxies and
spatially correlated with the lens sample.  At $R_{\rm eq}$ at
$z=0.67$, we show that almost all blends occur with the lens sample with
$M_{*}=10^{10.3}$\msun whereas only about $30$ per cent blends occur with
lens galaxies with $M_{*}=10^{11.0}$\msun.

Using a simple criterion based on the overlap between Kron
ellipses, we study how $N_{\rm pairs}(r)$ varies with $f$.  Over all
the stellar mass range, larger $f$ rejects more source galaxies and
results in lower \sn, which modifies the amplitude of \sn 
a factor less than $2$ from $f=0.8$ to $1.3$ and does not
change the peak of \sn as a function of the stellar mass.

Finally, we use our COSMOS weak lensing catalogue to make a first
prediction for the \sn for weak lensing measurements at $r<R_{\rm eq}$
for {\it Euclid} and {\it WFIRST}. Our predictions show that future experiments
should have enough source galaxies on small scales to detect weak
lensing down to a few tens of kpc with high \sn. Our
predictions show that this idea is worth pursuing further - but make a
number of simplifying assumptions that need to be investigated in
further detail.

Our work has focused primarily on quantifying the effects on blends on
the number of source galaxies at $r<R_{eq}$. In terms of raw numbers,
our work shows that even after making a fairly conservative selection
for our source galaxies, {\it Euclid} and {\it WFIRST} should still
have a sufficient number of pairs on small scales to measure weak
lensing with high \sn. This type of measurements however, will face a
number of challenges which still need to be investigated. The main
challenges that remain to be tackled will be understanding how to
measure both shear and photometric redshifts in an unbiased fashion in
close-pair configurations. In addition, further work is needed to test
and calibrate shear measurements in this intermediate regime where the
shear takes on values above 0.05. Also, here we have only considered
stellar mass, but in addition the gas component may also be
non-negligible especially at higher redshifts
\citep[]{Papastergis2012}. Finally, but not least, understanding how
to separate source galaxies from other galaxies spatially associated
with lens galaxies and how to correct for boost factors
\citep[][]{Mandelbaum2005} represent another challenge to be
investigated in greater detail.

\section*{Acknowledgements}

We are grateful to the referee for a careful reading of the manuscript and
for providing thoughtful comments.
We thank Robert Lupton for useful discussions during the preparation
of this paper and Naoshi Sugiyama for practical advice during the data analysis.
This work, AL, and SM are supported by World Premier International
Research Center Initiative (WPI Initiative), MEXT, Japan.
MINK acknowledges the financial support from N. Sugiyama (25287057)
by Grants-in-Aid from the Ministry of Education, Culture,
Sports, Science, and Technology of Japan.
N. Okabe(26800097) is supported by Grants-in-Aid from the Ministry of Education, Culture,
Sports, Science, and Technology of Japan.
C. Laigle is supported by the ILP LABEX (under reference ANR-10-LABX-63 and ANR-11-IDEX-0004-02).
J. Rhodes was supported by JPL, run under a contract for NASA by Caltech.
TTT has been supported by the Grant-in-Aid for the Scientific
Research Fund (23340046), for the Global COE Program Request
for Fundamental Principles in the Universe: from Particles to
the Solar System and the Cosmos, and for the JSPS Strategic
Young Researcher Overseas Visits Program for Accelerating
Brain Circulation, commissioned by the Ministry of Education,
Culture, Sports, Science and Technology (MEXT) of Japan.


\bibliographystyle{mn2e}
\bibliography{galaxy}


\end{document}